\documentclass[paper=a4, fontsize=11pt]{scrartcl} 

\usepackage{docmute}
\usepackage[sort&compress,numbers]{natbib}
\usepackage{subfig}
\usepackage{graphicx}
\usepackage{fancybox}
\usepackage{algorithm,algorithmic}
\usepackage[mode=buildnew]{standalone}
\usepackage{tikz,pgfplots,subfig}
\usepackage{multirow}
\usepackage{wrapfig}
\usepackage{grffile}
\pgfplotsset{compat=newest}
\usetikzlibrary{plotmarks}
\usetikzlibrary{positioning,arrows}

\usepgfplotslibrary{groupplots}
\usepackage{amsmath}
\usepackage{enumitem}

\usetikzlibrary{plotmarks} 

\usepackage[T1]{fontenc} 
\usepackage{fourier} 
\usepackage[english]{babel} 
\usepackage{amsmath,amsfonts,amsthm,amssymb} 
\usepackage{theorem}

\usepackage{lipsum} 

\usepackage{sectsty} 
\allsectionsfont{\centering \normalfont\scshape} 

\usepackage{fancyhdr} 
\numberwithin{equation}{section} 
\numberwithin{figure}{section} 
\numberwithin{table}{section} 

\newcommand{\horrule}[1]{\rule{\linewidth}{#1}} 

\title{
\normalfont \normalsize 
\textsc{ENSEEIHT} \\ [25pt] 
\horrule{0.5pt} \\[0.4cm] 
\huge Bayesian Structured Sparsity Priors for EEG Source Localization Technical Report \footnote{An extended version of a paper submitted for publication: \textit{Bayesian Structured Sparsity Priors for EEG Source Localization.}}\\ 
\horrule{2pt} \\[0.5cm] 
}

\author{Facundo~Costa, Hadj~Batatia, Thomas~Oberlin, and Jean-Yves~Tourneret} 

\date{\normalsize\today} 

\def\PM{\kern0pt^{\textrm{{\scriptsize PM}}}\kern0pt}
\def\MMAP{\kern1pt^{\textrm{{\tiny MMAP}}}\kern-1pt}

\def\MAP{^{\textrm{{\tiny MAP}}}}

  \def\cb{{\sbm{c}}\XS}
\def\Db{{\sbm{D}}\XS}  
\def\Eb{{\sbm{E}}\XS}

\def\Hb{{\sbm{H}}\XS}  \def\hb{{\sbm{h}}\XS}
\def\Ib{{\sbm{I}}\XS}

\def\Mb{{\sbm{M}}\XS}  \def\mb{{\sbm{m}}\XS}

  \def\pb{{\sbm{p}}\XS}
\def\Qb{{\sbm{Q}}\XS}  
\def\Rb{{\sbm{R}}\XS}  \def\rb{{\sbm{r}}\XS}
  \def\sb{{\sbm{s}}\XS}

  \def\vb{{\sbm{v}}\XS}
  
\def\Xb{{\sbm{X}}\XS}  \def\xb{{\sbm{x}}\XS}
\def\Yb{{\sbm{Y}}\XS}  \def\yb{{\sbm{y}}\XS}
  \def\zb{{\sbm{z}}\XS}
\def\mub{{\sbm{\mu}}\XS}

 \def\+{^\dagger}

\usepackage{afterpage}
\usepackage{textcomp}
\RequirePackage{xspace}

\newcommand{\diag}{\ensuremath{\mathrm{diag}}}

\newcommand{\RR}{\ensuremath{\mathbb R}}

\theoremstyle{plain}{\theorembodyfont{\rmfamily}%
\theoremstyle{plain}{\theorembodyfont{\rmfamily}%

\def\qed{\ifmmode\hbox{\hfill\sqb}\else{\ifhmode\unskip\fi%
\nobreak\hfil
\penalty50\hskip1em\null\nobreak\hfil$\blacksquare$
\parfillskip=0pt\finalhyphendemerits=0\endgraf}\fi}

\def\argmax{\mathop{\mathrm{arg\,max}}} 
\RequirePackage{amsmath}
\RequirePackage{xspace}
\RequirePackage{bbm}

\def\XS{\xspace}
\DeclareMathAlphabet{\mathb}{OML}{cmm}{b}{it}
\def\sbm#1{\ensuremath{\mathb{#1}}}                
\def\sbmm#1{\ensuremath{\boldsymbol{#1}}}          

  \def\cb{{\sbm{c}}\XS}
\def\Db{{\sbm{D}}\XS}  
\def\Eb{{\sbm{E}}\XS}

\def\Hb{{\sbm{H}}\XS}  \def\hb{{\sbm{h}}\XS}
\def\Ib{{\sbm{I}}\XS}

\def\Mb{{\sbm{M}}\XS}  
\def\mb{{\sbm{m}}\XS}

  \def\pb{{\sbm{p}}\XS}
\def\Qb{{\sbm{Q}}\XS}  
\def\Rb{{\sbm{R}}\XS}  \def\rb{{\sbm{r}}\XS}
  \def\sb{{\sbm{s}}\XS}

  \def\vb{{\sbm{v}}\XS}
  
\def\Xb{{\sbm{X}}\XS}  \def\xb{{\sbm{x}}\XS}
\def\Yb{{\sbm{Y}}\XS}  \def\yb{{\sbm{y}}\XS}
  \def\zb{{\sbm{z}}\XS}

\def\Hb{{\sbm{H}}\XS}  \def\hb{{\sbm{h}}\XS}

 \def\+{^\dagger}

\def\MAP{^{\kern1pt{\rm MAP}\kern-1pt}}

\usepackage{color,soul}

\def\mub         {{\sbmm{\mu}}\XS}

\def\taub        {{\sbmm{\tau}}\XS}

\def\PM{\kern0pt^{\textrm{{\scriptsize PM}}}\kern0pt}
\def\MMAP{\kern1pt^{\textrm{{\tiny MMAP}}}\kern-1pt} 

\def\rem#1{}                    

\begin{document}

\maketitle 
\begin{center}
\textbf{Abstract}
\end{center}

\begin{abstract}
This report introduces a new hierarchical Bayesian model for the EEG source localization problem. This model promotes structured sparsity to search for focal brain activity. This sparsity is obtained via a multivariate Bernoulli Laplacian prior assigned to the brain activity approximating an $\ell_{20}$ pseudo norm regularization in a Bayesian framework. A partially collapsed Gibbs sampler is used to draw samples asymptotically distributed according to the posterior associated with the proposed Bayesian model. The generated samples are used to estimate the brain activity and the model hyperparameters jointly in an unsupervised framework. Two different kinds of Metropolis-Hastings moves are introduced to accelerate the convergence of the Gibbs sampler. The first move is based on multiple dipole shifts within each MCMC chain whereas the second one exploits proposals associated with different MCMC chains. We use both synthetic and real data to compare the performance of the proposed method with the weighted $\ell_{21}$ mixed norm regularization and a method based on a multiple sparse prior, showing that our algorithm presents advantages in several scenarios.
\end{abstract}

\section{Introduction}
\label{sec:intro}
The EEG source localization problem continues to attract a high level of coverage in the literature resulting in a wide array of methods developed in the last years. These can be classified in two groups: (i) the dipole-fitting models that represent the brain activity as a small number of dipoles with unknown positions; and (ii) the distributed-source models that represent it as a large number of dipoles in fixed positions. Dipole-fitting models \cite{Buchner1997,cuffin1995method} try to estimate the amplitudes, orientations and positions of a few dipoles that explain the measured data. Unfortunately, these models usually provide solutions that vary with the initial guess of the number of dipoles and with their initial locations due to the presence of a large number of local minima in their cost function \cite{grech2008review}. Several algorithms based on MUSIC were developed to solve this problem \cite{mosher1992multiple,mosher1998recursive,mosher1999source,xu2004alternative}. In addition, sequential Monte Carlo methods were also investigated to estimate the dipole-fitting model parameters \cite{sommariva2014sequential}. If the brain activity is composed of a small number of clustered sources, the dipole-fitting algorithms are capable of providing good results \cite{da2005biophysical,liu2005standardized}. However, their performance deteriorates for detecting multiple spatially extended sources \cite{grech2008review}. On the other hand, the distributed-source methods model the brain activity as the result of a large number of discrete dipoles with fixed positions and try to estimate their amplitudes and orientations \cite{grech2008review}. Since the amount of dipoles used in the brain model is typically much larger than the amount of electrodes, the inverse problem is ill-posed in the sense that there is an infinite amount of brain activities that can justify the measurements \cite{grech2008review}. A regularization is thus needed in order to incorporate additional information to solve this inverse problem. One of the most simple regularizations consists of penalizing the $\ell_2$ norm of the solution using the minimum norm estimation algorithm \cite{pascual1999review} or its variants based on the weighted minimum norm: Loreta \cite{pascual1994low} and sLoreta \cite{pascual2002standardized}. However, these methods have been shown to overestimate the size of the active area if the brain activity is focused \cite{grech2008review}, which is believed to be the case in a number of medical applications.  
A better way to estimate focal brain activity is to promote sparsity, by applying an $\ell_0$ pseudo norm regularization \cite{candes2008restricted}. Unfortunately, this procedure is known to be intractable in an optimization framework. As a consequence, the $\ell_0$ pseudo norm is usually approximated by the $\ell_1$ norm via convex relaxation \cite{uutela1999visualization}, in spite of the fact that these two approaches do not always provide the same solution \cite{candes2008restricted}. In a previous work, we proposed to combine them in a Bayesian framework \cite{costa2015bayesl0l1}, using the $\ell_0$ penalty to locate the non-zero positions and the $\ell_1$ norm to estimate their amplitudes. However, this $\ell_0$ + $\ell_1$ method, similarly to $\ell_0$ and $\ell_1$ separately, considers each time sample independently leading in some cases to unrealistic solutions \cite{gramfort2012mixed}.

To improve source localization, it is possible to make use of the temporal structure of the data by promoting structured sparsity, which is known to yield better results than standard sparsity when applied to strongly group sparse signals \cite{huang2010benefit}. Structured sparsity has been shown to improve results in several applications including audio restoration \cite{kowalski2013social}, image analysis \cite{yu2012solving} and machine learning \cite{huang2011learning}. One way of applying structured sparsity in EEG source localization is to use mixed-norms regularization such as the $\ell_{21}$ mixed norm \cite{gramfort2012mixed} (also referred to as group-lasso). This approach promotes sparsity among different dipoles (via the $\ell_1$ portion of the norm) but  groups all the time samples of the same dipole together, forcing them all to be either jointly active or inactive (with the $\ell_2$ norm portion). However, it has several drawbacks including  the manual tuning of the regularization parameter.

In addition to optimization techniques, several approaches have tried to model the time evolution of the dipole activity and estimate it using either Kalman filtering \cite{galka2004solution,long2011state} or particle filters \cite{somersalo2003non,sorrentino2013dynamic,chen2013multiple}. Several Bayesian methods have been used as well, both in dipole-fitting models \cite{kiebel2008variational,jun2005spatiotemporal} and distributed source models \cite{friston2008multiple,stahlhut2013hierarchical}. In \cite{friston2008multiple}, Friston \textit{et al.} developed the multiple sparse priors (MSP) approach, in which they parcellate the brain in different pre-defined regions and promote all the dipoles in each region to be active or inactive jointly. Doing this they encourage the brain activity to extend over an area instead of being focused in point-like sources. Conversely, we are mainly interested in estimating point-like focal source activity which has been proved to be relevant in clinical applications \cite{huppertz2001cortical}. In order to do this, we will consider each dipole separately instead of grouping them together. Note that this approach avoids the need of choosing a criterion for brain parcellization as required in the MSP method.

This report introduces a new hierarchical Bayesian model that estimates the brain activity with a structured sparsity constraint by using a multivariate Bernoulli Laplace prior (approximating the weighted $\ell_{20}$ mixed norm). Since the parameters of the proposed model cannot be computed with closed-form expressions, we investigate Markov chain Monte Carlo sampling techniques to draw samples that are asymptotically distributed according to the posterior of the proposed model. We then estimate jointly the brain activity, the model parameters and hyperparameters in an unsupervised framework. In order to avoid the sampler to get stuck around local maxima, specific Metropolis-Hastings moves are introduced, allowing new modes of the posterior to be explored. These moves are based on multiple dipole shifts (moving active dipoles to neighboring positions) and inter-chain proposals (exchanging samples between parallel MCMC chains) that significantly accelerate the convergence speed of the proposed sampler. These proposals generate candidates that are accepted or rejected using a Metropolis-Hastings criterion. The method is applied to both synthetic and real data showing promising results compared to the more traditional $\ell_{21}$ mixed norm and the MSP method.

The report is organized as follows: Section \ref{sec:prob_statement} describes the considered problem. The proposed Bayesian model is presented in Section \ref{sec:bayesian_model}. Section \ref{sec:gibbs_sampler} introduces the partially collapsed Gibbs sampler used to generate samples distributed according to the posterior of this model. The Metropolis-Hastings moves that are used to accelerate the convergence of the sampler are introduced in Section \ref{sec:convergence_considerations}. Experimental results conducted for both synthetic and real data are presented in Section \ref{sec:exp_results}.
Conclusions are finally reported in Section \ref{sec:conclusion}.
Appendices \ref{appendix:cond_dist_derivation} and \ref{appendix:accept_proposal_prob} include the algebraic derivations of the conditional distributions used in the Gibbs sampler and the acceptance rate of the Metropolis-Hastings moves respectively.

\section{Problem statement}
\label{sec:prob_statement}
The EEG source localization is an inverse problem that consists in estimating the brain activity of a patient from EEG measurements taken from $M$ electrodes during $T$ time samples. In a distributed source model, the brain activity is represented by a finite number of dipoles located at fixed positions in the brain cortex. More precisely, we consider $N$ dipoles located in the surface of the brain cortex and oriented orthogonally to it (motivations for this can be found in \cite{hallez2007review}). The EEG measurement matrix $\Yb \in \RR^{M\times T}$ can be written
\begin{equation}
\Yb = \Hb \Xb + \Eb
\end{equation}
where $\Xb \in \RR^{N\times T}$ contains the dipole amplitudes, $\Hb \in \RR^{M\times N}$ represents the head operator (usually called ``leadfield matrix''),  and $\Eb$ is the measurement noise. Denote as $\mb_i$ the $i$-th row of the matrix $\Mb$ and as $\mb^j$ its $j$-th column.
 
Thus, the EEG source localization problem consists in estimating the matrix $\Xb$ from the known operator $\Hb$ and the measurements $\Yb$.

The next section introduces the hierarchical Bayesian method proposed to solve this inverse problem.

\section{Bayesian model}
\label{sec:bayesian_model}

\subsection{Likelihood}
As is classical in the literature, we consider an additive white Gaussian noise with a constant variance $\sigma_n^2$ over the considered time samples \cite{grech2008review}. Note that when this assumption does not hold it is possible to estimate the noise covariance matrix from measurements that do not contain the signal of interest and use it to whiten the data \cite{maris2003resampling}. This leads to a gaussian likelihood

\begin{equation}
\label{eq:l21_likelihood}
f(\Yb | \Xb, \sigma_n^2) = \prod_{t=1}^T \mathcal{N}\Big(\yb^t \Big| \Hb \xb^t, \sigma_n^2 I_M\Big)
\end{equation}
where $I_M$ is the identity matrix of size $M \times M$.

\subsection{Prior distributions}
\subsubsection{Prior of the brain activity $\Xb$}
To promote structured sparsity of the source activity, we first consider the weighted $\ell_{20}$ mixed pseudo-norm

\begin{equation}
||\Xb||_{20} = \#\{i : \sqrt{v_i} ||\xb_i||_2 \neq 0\}
\end{equation}
where $v_i = ||\hb^i||_2$ is a weight introduced to compensate the depth-weighting effect \cite{grech2008review,uutela1999visualization} and $\#\mathcal{S}$ denotes the cardinal of the set $\mathcal{S}$. Since this prior leads to intractable computations, we propose to approximate it by a multivariate Laplace Bernoulli prior for each row of $\Xb$

\begin{equation}
f(\xb_i | z_i, \lambda) \propto 
\left\{
	\begin{array}{ll}
		\delta(\xb_i)  & \mbox{if } z_i = 0 \\
		\exp\Big(-\frac{1} {\lambda}\sqrt{v_i}||\xb_i||_2\Big) & \mbox{if } z_i = 1
	\end{array}
\right.
\end{equation}

where $\lambda$ is the exponential distribution parameter and $\zb \in \{0, 1\}^{N}$ is a vector indicating if the rows of $\Xb$ are non-zero. To make the analysis easier it is convenient to define the hyperparameter $a = \frac{\sigma_n^2} {\lambda^2}$ which transforms the prior to

\begin{equation}
\label{eq:prior_x}
f(\xb_i | z_i, a, \sigma_n^2) \propto
\left\{
	\begin{array}{ll}
		\delta(\xb_i)  & \mbox{if } z_i = 0 \\
		\exp\Big(-\sqrt{\frac{v_i a} {\sigma_n^2}} ||\xb_i||_2\Big) & \mbox{if } z_i = 1
	\end{array}
\right.
\end{equation}

The elements $z_i$ are assigned a Bernoulli prior distribution with parameter $\omega \in [0, 1]$:

\begin{equation}
f(z_i | \omega) = \mathcal{B}\Big(z_i | \omega\Big)
\end{equation}

Note that the Dirac delta function $\delta(.)$ in the prior of $\xb_i$ promotes sparsity while the Laplace distribution regulates the amplitudes of the non-zero rows. The parameter $\omega$ tunes the balance between them. Indeed, $\omega = 0$ yields $\Xb = 0$ whereas $\omega = 1$ reduces the prior to the Bayesian formulation of the group-lasso \cite{yuan2006model}. Unfortunately the prior \eqref{eq:prior_x} leads to an intractable posterior. It is possible to fix this problem by introducing a latent variable vector $\taub^2 \in (\RR^+)^N$ as proposed in \cite{raman2009bayesian}. More precisely, we use the following gamma and Bernoulli-Gaussian priors on $\tau^2_i$ and $\xb_i$ respectively

\begin{equation}
f(\tau_i^2 | a) = \mathcal{G}\Big(\tau_i^2 \Big| \frac{T + 1} {2}, \frac{v_i a} {2}\Big)
\end{equation}

\begin{equation}
f(\xb_i | z_i, \tau_i^2, \sigma_n^2) = 
\left\{
	\begin{array}{ll}
		\delta(\xb_i)  & \mbox{if } z_i = 0 \\
		\mathcal{N}\Big(\xb_i \Big| 0, \sigma_n^2 \tau_i^2 I_T\Big) & \mbox{if } z_i = 1
	\end{array}
\right.
\end{equation}

which can be shown to lead to the desired marginal distribution of $\xb_i$ \eqref{eq:prior_x} \cite{raman2009bayesian}.

In addition assuming the rows of $\taub^2$, $\zb$, and $\Xb$ apriori independent leads to the following priors:

\begin{align*}
f(\taub^2 | a) &= \prod_i^N {f(\tau^2_i | a)} \\
f(\zb | \omega) &= \prod_i^N {f(z_i | \omega)} \\
f(\Xb | \zb, \taub^2, \sigma_n^2) &= \prod_i^N {f(\xb_i | z_i, \tau^2_i, \sigma_n^2)}
\end{align*}

\subsubsection{Prior of the noise variance activity $\sigma_n^2$}
The noise variance is assigned a Jeffrey's prior

\begin{equation}
\label{eq:21_jeffrey_prior_sigma}
f(\sigma_n^2) \propto \frac{1} {\sigma_n^2} 1_{\RR^+} (\sigma_n^2)
\end{equation}
where $1_{\RR^+}(\xi)=1$ if $\xi \in \RR^+$ and 0 otherwise. This choice is very classical when no information about a scale parameter is available (see \cite{casella1999monte} for details).

\subsection{Hyperparameter priors}
\begin{figure}[H]
\begin{center}
\begin{tikzpicture}[]
   \begin{scope}[
		node distance=1.6cm,on grid,>=stealth',
		hyperpar/.style={rectangle,draw},
		par/.style={circle,draw}]
   \node [par] 	(y)					{$\Yb$};
   \node [par]	(xs)[below=of y]		{$\Xb$, $\sigma_n^2$}  edge [->] (y);
   \node [par]	(z)	[below=of xs,xshift=0.8cm]	{$\zb$} edge [->] (xs);
   \node [par]	(t) [left=of z]		{$\taub^2$} edge [->] (xs);
   \node [hyperpar]	(a) [below=of t]	{$a$} edge [->] (t);
   \node [hyperpar]	(w) [below=of z]	{$\omega$} edge [->] (z);
   \end{scope}
\end{tikzpicture}
\end{center}   
\caption{Directed acyclic graph for the proposed Bayesian model.}
\label{fig:bayesian_herarchy}
\end{figure}
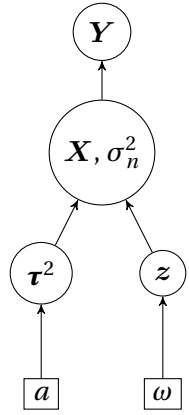

The proposed method allows one to balance the importance between sparsity of the solution and fidelity to the measurements using two hyperparameters: 1) $\omega$ that adjusts the proportion of non-zero rows, and 2) $a$ that controls the amplitudes of the non-zeros. The corresponding hierarchy of parameters and hyperparameters is shown in Fig. \ref{fig:bayesian_herarchy}. In contrast to the $\ell_{21}$ mixed norm our algorithm does not require to adjust these hyperparameters but is able to estimate their values from the data by assigning hyperpriors to them following a so-called hierarchical Bayesian analysis. This section defines the priors assigned to the model hyperparameters.

\subsubsection{Hyperprior of $a$}
A conjugate gamma prior is assigned to $a$
\begin{equation}
\label{eq:21_jeffrey_prior_a}
f(a | \alpha, \beta) = \mathcal{G}\Big(a \Big| \alpha, \beta\Big) 
\end{equation}
with $\alpha = \beta = 1$. These values of $\alpha$ and $\beta$ yield a vague hyperprior for $a$. The conjugacy of this hyperprior will make the analysis easier in the sense that the conditional distribution of $a$ required in the Gibbs sampler will also be a gamma distribution.

\subsubsection{Hyperprior of $\omega$}
A uniform prior on [0, 1] is used for $\omega$

\begin{equation}
f(\omega) = \mathcal{U}\Big(\omega \Big| 0, 1\Big)
\end{equation}
reflecting the absence of knowledge for this hyperparameter.

\subsection{Posterior distribution}
Using the previously described priors and hyperpriors, the posterior distribution of the proposed Bayesian model is
\begin{equation}
\label{eq:posterior}
f(\Yb, \sigma_n^2, \Xb, \zb, a, \taub^2, \omega) \propto f(\Yb | \Xb, \sigma_n^2) f(\Xb | \taub^2, \zb, \sigma_n^2) f(\zb | \omega) f(\taub^2 | a) f(\sigma_n^2) f(a) f(\omega)
\end{equation}
The following section investigates a partially collapsed Gibbs sampler that is used to sample according to the posterior distribution \eqref{eq:posterior} and to build estimators of the unknown model parameters and hyperparameters using these generated samples.

\section{A Partially collapsed Gibbs Sampler}
\label{sec:gibbs_sampler}
The posterior distribution \eqref{eq:posterior} is intractable and does not allow us to derive closed-form expressions for the estimators of the different parameters and hyperparameters of the proposed model. Thus we propose to draw samples from \eqref{eq:posterior} and to use them to estimate the brain activity jointly with the model hyperparameters. More precisely, we investigate a partially collapsed Gibbs sampler that samples the variables $z_i$ and $\xb_i$ jointly in order to exploit the strong correlation between these two variables. If we denote by $\Xb_{-i}$ the matrix $\Xb$ whose $i$th row has been replaced by zeros, the resulting sampling strategy is summarized in Algorithm \ref{algo:gibbs_sampler}. The corresponding conditional distributions are shown hereafter and their exact derivation can be found in Appendix \ref{appendix:cond_dist_derivation}.

\begin{algorithm}
\caption{Partially Collapsed Gibbs sampler.}
\begin{algorithmic}[]\label{algo:gibbs_sampler}
\STATE  Initialize $\Xb = 0$ and $\zb = 0$
\STATE  Sample $a$ and $\taub^2$ from their prior distributions

\REPEAT
  \STATE   Sample $\sigma_n^2$ from $f(\sigma_n^2 | \Yb, \Xb, \taub^2, \zb)$
  \STATE   Sample $\omega$ from $f(\omega | \zb)$
  \FOR {$i=1$ to $N$}
  	\STATE   Sample $\tau_i^2$ from $f(\tau_i^2 | \xb_i, \sigma_n^2, a, z_i)$
  	\STATE   Sample $z_i$ from $f(z_i | \Yb, \Xb_{-i}, \sigma_n^2, \tau_i^2, \omega)$
  	\STATE   Sample $\xb_i$ from $f(\xb_i | z_i, \Yb, \Xb_{-i}, \sigma_n^2, \tau_i^2)$
  \ENDFOR
  \STATE   Sample $a$ from $f(a | \taub^2)$
  \UNTIL {convergence}
\end{algorithmic}
\end{algorithm}

\subsubsection{Conditional distribution of $\tau_i^2$}
The conditional distribution of $\tau_i^2$ is a gamma distribution or a generalized inverse Gaussian distribution depending on the value of $z_i$. More precisely

\begin{equation}
f(\tau_i^2 | \xb_i, \sigma_n^2, a, z_i) =
\left\{
	\begin{array}{ll}
	\mathcal{G}\Big(\tau_i^2 \Big| \frac{T + 1} {2}, \frac{v_i a} {2}\Big) & \mbox{if } z_i = 0 \\
	\mathcal{GIG}(\tau_i^2 \Big| \frac{1} {2},  v_i a, \frac{||\xb_i||^2} {\sigma_n^2}\Big) & \mbox{if } z_i = 1.\\
	\end{array}
\right.
\end{equation}

\subsubsection{Conditional distribution of $\xb_i$}
The conditional distribution of the $i$th row of $\Xb$ is
\begin{equation}
f(\xb_i | z_i, \Yb, \Xb_{-i}, \sigma_n^2, \tau_i^2) = 
\left\{
	\begin{array}{ll}
		\delta(\xb_i)  & \mbox{if } z_i = 0 \\
		\mathcal{N}\Big(\xb_i \Big| \mub_i, \sigma_i^2\Big) & \mbox{if } z_i = 1.
	\end{array}
\right.
\end{equation}
with
\begin{align}
\mub_i = \frac{\sigma_i^2 ({\hb^i})^T (\Yb - \Hb \Xb_{-i})} {\sigma_n^2}, 
\sigma_i^2 = \frac{\sigma_n^2 \tau_i^2} {1 + \tau_i^2 ({\hb^i})^T \hb^i}
\end{align}

\subsubsection{Conditional distribution of $z_i$}
The conditional distribution of $z_i$ is a Bernoulli distribution
\begin{equation}
f(z_i | \Yb, \Xb_{-i}, \sigma_n^2, \tau_i^2, \omega) = \mathcal{B} \Big(z_i \Big| 1, \frac{k_1} {k_0 + k_1}\Big)
\end{equation}
with
\begin{align}
k_0 = 1 - \omega, k_1 = \omega {(\frac{\sigma_n^2 \tau_i^2} {\sigma_i^2})}^{-\frac{T} {2}}  \exp{\Big(\frac{||\mub_i||^2} {2 \sigma_i^2}\Big)}.
\end{align}

\subsubsection{Conditional distribution of $a$}
The conditional distribution of $a | \taub^2$ is the following gamma distribution

\begin{equation}
f(a | \taub^2) = \mathcal{G}\Big(a \Big| \frac{N(T+1)} {2} + \alpha, \frac{\sum_i[v_i \tau_i^2]} {2} + \beta\Big).
\end{equation}

\subsubsection{Conditional distribution of $\sigma_n^2$}
The distribution of $\sigma_n^2 | \Yb, \Xb, \taub^2, \zb$ is the following inverse gamma distribution

\begin{equation}
f(\sigma_n^2 | \Yb, \Xb, \taub^2, \zb) = \mathcal{IG}\Big(\sigma_n^2 \Big| \frac{(M+||\zb||_0)T} {2}, \frac{1} {2} \Big[||\Hb \Xb - \Yb||^2 + \sum_{i \in I_1} \frac{||\xb_i||^2} { \tau_i^2}\Big]\Big).
\end{equation}

\subsubsection{Conditional distribution of $\omega$}
Finally, $\omega | \zb$ has the following beta distribution

\begin{equation}
f(\omega | \zb) = \mathcal{B}e\Big(\omega \Big| 1 + ||\zb||_0, 1 + N - ||\zb||_0\Big).
\end{equation}

\begin{figure}
\centering
\includegraphics[scale = 0.6]{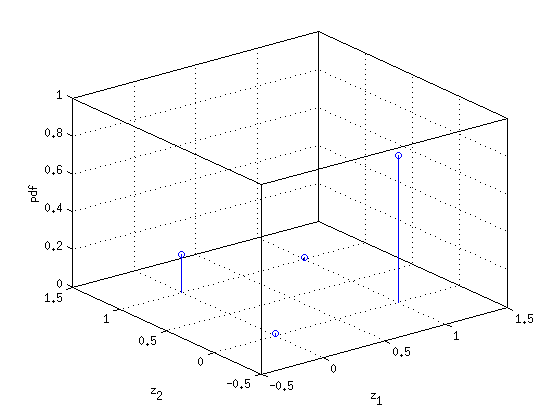}
\caption{Example of posterior distribution of $\zb$ with local maxima}
\label{fig:local_maxima_z}
\end{figure}

\begin{figure}[]
	\centering
	\subfloat[][Ground truth - Axial, coronal and sagittal views respectively]{
		\includegraphics{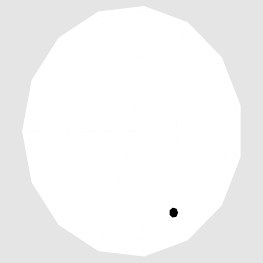}
		\includegraphics{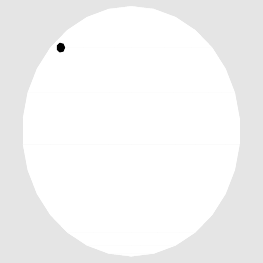}
 		\includegraphics{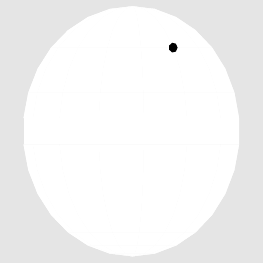}
	}

	\centering
	\subfloat[][Estimation without proposals - Axial, coronal and sagittal views respectively]{
		\includegraphics{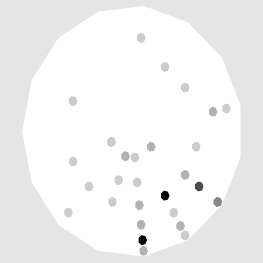}
		\includegraphics{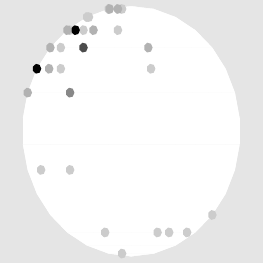}
 		\includegraphics{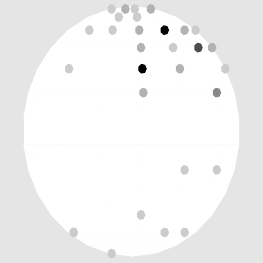}
	}

	\subfloat[][Estimation with proposals - Axial, coronal and sagittal views respectively]{
		\includegraphics{Figs/proposals_illustrating/intra_with_location_ax.pdf}
		\includegraphics{Figs/proposals_illustrating/intra_with_location_cor.pdf}
 		\includegraphics{Figs/proposals_illustrating/intra_with_location_sag.pdf}
	}
		
	\caption{Illustration of the effectiveness of the multiple dipole shift proposals.}
	\label{fig:proposals_intra_prob_detection}
\end{figure}

\section{Convergence considerations}
\label{sec:convergence_considerations}
\subsection{Local maxima}
We have observed that the proposed partially collapsed Gibbs sampler may be stuck around local maxima of the variable $\zb$ from which it is very difficult to escape in a reasonable amount of iterations. Fig. \ref{fig:local_maxima_z} illustrates via a simple example that if the sampler gets to $\zb = \{0, 1\}$ it requires to go through intermediary states with very low probability to move to the correct value $\zb = \{1, 0\}$. This kind of situation can occur if the dipoles corresponding to $z_1$ and $z_2$ produce similar measurements $\Yb$ when active so that the probability of having either of them active is much higher than having them both on (or off) at the same time. 


\subsection{Multiple dipole shift proposals}
In order to solve this problem, we introduce a proposal making scheme, that consists in changing several elements of $\zb$ simultaneously (which would allows us to go from $\{0, 1\}$ to $\{1, 0\}$ in one step in the previous example) after each sampling iteration. The proposal is accepted or rejected using a Metropolis-Hastings criteria which guarantees that the target distribution is preserved.

We have observed that one of the most common ways for the algorithm to get stuck in local maxima is by failing to estimate the position of the non-zero elements of $\zb$. In other words, the sampling scheme detects a non-zero in a position that is not correct but is close (in the sense that it produces similar measurements) to the correct one, which causes the problem described above.

\subsubsection{Shift based proposals}
Before describing the proposed move, it is interesting to mention that it was inspired by an idea developed by Bourguignon \textit{et al} in \cite{bourguignon2005bernoulli} to perform a spectral analysis of astrophysical data obtained after irregular sampling. The authors of \cite{bourguignon2005bernoulli} proposed to move a single non-zero element of a binary sequence to a random neighboring position after each iteration of the MCMC sampler. In the presence of a single non-zero element, this move is sufficient to escape from a local maximum of the posterior associated with our EEG source localization model. However, when there are several non-zero elements located at wrong locations, proposing to move each one of them separately may not be sufficient to escape from a local maximum. For this reason, we have generalized the scheme presented in \cite{bourguignon2005bernoulli} by proposing to move a random subset of $K$ estimated non-zeros simultaneously to random neighboring positions. According to our experiments (some of them described in Section \ref{sec:exp_results}), the simple choice $K = 2$ provides good results in most practical cases. Since there is a high correlation between the variables $\taub^2$ and $\zb$, it is convenient to update their values jointly. The resulting multiple dipole shift proposal is detailed in Algorithm \ref{algo:proposal}.
Note that Bourguignon \textit{et al} work with 1-dimensional data so they define the neighborhood of the element $z_k$ as  $\{z_{k-1}, z_{k+1}\}$. In contrast, we are working with dipoles located in a 3-dimensional brain so the neighborhood definition is non-trivial and will be described in the following.

\begin{algorithm}
\caption{Multiple dipole shift proposal.}
\begin{algorithmic}[!H]\label{algo:proposal}
\STATE   $\bar{\zb} = \zb$
  \STATE \textbf{repeat} K times
  	\STATE\hspace{\algorithmicindent}	Set $\textrm{ind}_{\textrm{old}}$ to be the index of a random non-zero of $\zb$
  	\STATE\hspace{\algorithmicindent}	Set $\pb = [\textrm{ind}_{\textrm{old}}, \textrm{neigh}_{\gamma}(\textrm{ind}_{\textrm{old}})]$
  	\STATE\hspace{\algorithmicindent} 	Set $\textrm{ind}_{\textrm{new}}$ to be a random element of $\pb$ 
  	\STATE\hspace{\algorithmicindent}	Set $\bar{z}_{\textrm{ind}_{\textrm{old}}} = 0$ and $\bar{z}_{\textrm{ind}_{\textrm{new}}} = 1$
  \STATE \textbf{end}
  	\STATE	Sample $\bar{\Xb}$ from $f(\bar{\Xb} | \bar{\zb}, \Yb, \sigma_n^2, \taub^2)$.    
  	\STATE	Sample $\bar{\taub}^2$ from $f(\bar{\taub}^2 | \bar{\Xb}, \sigma_n^2, a, \bar{\zb})$.
  \STATE		Set $\{\zb,\taub^2\} = \{\bar{\zb}, \bar{\taub}^2\}$ with probability $\min\Big(\frac {f(\bar{\zb}, \bar{\taub}^2 | .)} {f(\zb, \taub^2 | .)}, 1\Big)$
  \STATE		Resample $\Xb$ if the proposal was accepted
\end{algorithmic}
\end{algorithm}

In Fig. \ref{fig:proposals_intra_prob_detection} we can see the effect of introducing multiple dipole shift proposals (with $K = 2$) in a practical case. The first row of images is the ground truth (a single dipole activation) while the second row shows the probability of finding each dipole active with eight MCMC parallel chains without using proposals after 10.000 iterations. As we can see, the activity is in the correct area but the algorithm is not able to converge to the correct value of $\zb$. After introducing the multiple dipole shift proposals the sampler converges in less than 1.000 iterations as shown in the third row of the figure.

\subsubsection{Acceptance probability}
In order to guarantee that the generated samples are asymptotically distributed according to the posterior of the proposed model, all moves resulting from the multiple dipole shift proposal are accepted or rejected with the acceptance probability described in Algorithm \ref{algo:proposal} that uses the following probability distribution

\begin{align}
&f(\zb_r, \taub_r^2 | \Yb, a, \sigma_n^2, \omega) \propto (1-\omega)^{C_0} \omega^{C_1} (\sigma_n^2)^{-\frac{T C_1} {2}} |\Sigma|^{\frac{T} {2}} \\
&\prod_{i \in \Ib_1} {(\tau_i^2)^{-\frac{T} {2}}} \exp\Big({-\frac{\sum_{t=1}^T K^t} {2}}\Big) \prod_{i = 1}^{N}{\mathcal{G}\Big(\tau_i^2 \Big| \frac{T + 1} {2}, \frac{v_i a} {2}\Big)} \nonumber
\end{align}
where $\rb = \{i : z_i \neq \bar{z}_i \}$, $\Ib_k = \{i : z_{r_i} = k\}$, $C_k = \#\Ib_k$ for $k = \{0, 1\}$ and
\begin{align}
\Sigma^{-1} &= {\frac{1} {\sigma_n^2}\Big[{({\Hb^{\Ib_1}})^T \Hb^{\Ib_1} + \diag\Big(\frac{1} {\tau^2_{\rb}}\Big)}\Big]} \nonumber\\
\mu^t &= -\frac{\Sigma ({\Hb^{\Ib_1}})^T ({\Hb^{-\rb}} \xb_{-\rb}^t - \yb^t)} {\sigma_n^2} \nonumber\\
K^t &= \frac{(\Hb^{-\rb} \xb_{-\rb}^t - \yb^t)^T (\Hb^{-\rb} \xb_{-\rb}^t - \yb^t)} {\sigma_n^2} - ({\mu^t})^T  \Sigma^{-1} \mu^t. \nonumber
\end{align}
Note that $\mb_{-\sb}$ is obtained after removing in the vector $\mb$ all the rows belonging to the set $\sb$, $\Mb^{-\sb}$ is the matrix $\Mb$ whose columns belonging to $\sb$ have been removed, $\diag(\sb)$ is the diagonal square matrix whose diagonal elements are the elements of $\sb$ and $|\Mb|$ is the determinant of the matrix $\Mb$.
The exact derivation is included in Appendix \ref{appendix:accept_proposal_prob}.

\subsubsection{Neighborhood definition}
\label{sec:neighborhood}
It is obvious that the definition of the neighborhood used to exchange non-zero elements is crucial. Initially, we used a geometrical neighborhood, defined in terms of vertex connexity in the triangular tessellation modeling the brain cortex. However, this definition usually yields very small neighborhoods. This may cause the proposals not to be flexible enough to help the algorithm escape from local maxima.

For this reason we propose a neighborhood definition that considers two dipoles to be neighbors if the correlation between their respective columns is higher than a certain threshold

\begin{equation}
\label{eq:neigh_def}
\textrm{neigh}_{\gamma}(i) \triangleq \Big\{j \neq i \enspace \Big|\enspace |\textrm{corr}(\hb^i, \hb^j)| \geq \gamma\Big\}
\end{equation}

where $\textrm{corr}(\vb_1, \vb_2)$ is the correlation between vectors $\vb_1$ and $\vb_2$ and where the neighborhood size can be adjusted by setting $\gamma \in [0,1]$ ($\gamma = 0$ corresponds to a neighborhood containing all the dipoles and $\gamma = 1$ corresponds to an empty neighborhood).

An additional advantage of this definition is the fact that it allows the approach to be extended to other kind of inverse problems (different from EEG) where no geometrical disposition of the elements of $\zb$ may be available.

In order to maximize the moves efficiency, the value of $\gamma$ has to be selected carefully. A very large value of $\gamma$ will result in proposals not being flexible enough to help the algorithm in escaping local maxima. A very low value of $\gamma$ will result in a very large amount of possible proposals with many of them being useless leading to a large number of iterations to reach useful moves. Our experiments have shown that a good compromise is obtained with $\gamma = 0.8$ (see Section \ref{sec:exp_results} for illustrations).

\subsection{Inter-chain proposals}
The multiple dipole shift proposal scheme previously described allows the algorithm to better sample the value of $\zb$ present in the posterior distribution and is able to find the active dipoles correctly if they are few in number. However, when running multiple MCMC chains in parallel with a higher amount of non-zeros present in the ground truth, it is possible for the different chains to get stuck in different values of $\zb$. In order to help them converge to the same (most probable) value, it is possible to exchange information between parallel chains to avoid local maxima during their runs as other approaches do, including Metropolis-coupled MCMC \cite{geyer1991markov}, Population MCMC \cite{laskey2003population} and simulated tempering \cite{geyer1995annealing,marinari1992simulated}. 

In this report, we introduce inter-chain moves by proposing to exchange the values of $\zb$ and $\taub^2$ between different chains. This move is accepted with the Metropolis-Hastings probability shown in Algorithm \ref{algo:inter_proposal}.

One inconvenience introduced by inter-chain proposals is the fact that they require synchronizing the parallel MCMC chain processes, which decreases the iteration speed of the algorithm. In order to minimize this effect, an inter-chain proposal will be made after each iteration with probability $p$ (adjusted to $\frac{1} {1000}$ by cross validation) according to Algorithm \ref{algo:inter_proposal}.

\begin{algorithm}
\caption{Inter-chain proposals.}
\begin{algorithmic}[!H]\label{algo:inter_proposal}
\STATE   Define a vector $\cb = \{1, 2, ..., L\}$ where L is the number of chains
  \STATE \textbf{for} $i = \{1, 2, ..., L\}$
  	\STATE\hspace{\algorithmicindent}	Choose (and remove) a random element from $\cb$ and denote it by $k$
	\STATE\hspace{\algorithmicindent} Denote as $\{\bar{\zb}_k, \bar{\taub}_k^2\}$ the sampled values of $\{\zb, \taub^2\}$ of MCMC chain number $\#k$
	  \STATE	\hspace{\algorithmicindent} 	For the chain $\#i$ set $\{\zb_i,\taub_i^2\} = \{\bar{\zb}_k,\bar{\taub}_k^2\}$ with probability $\frac {f({\bar{\zb}_k}, \bar{\taub}_k^2 | .)} {f(\zb, \taub^2 | .)}$
  \STATE\hspace{\algorithmicindent}		Resample $\Xb$ if the proposal has been accepted
  \STATE \textbf{end}
\end{algorithmic}
\end{algorithm}

\begin{figure}[]
	\centering
	\subfloat[][Ground truth - Axial, coronal and sagittal views respectively]{
		\includegraphics{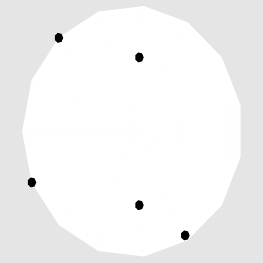}
		\includegraphics{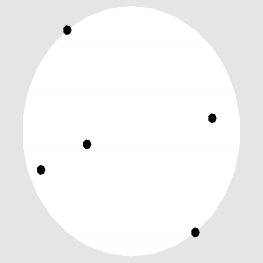}
 		\includegraphics{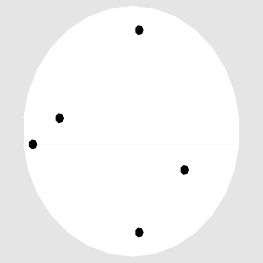}
	}

	\centering
	\subfloat[][Estimation without proposals - Axial, coronal and sagittal views respectively]{
		\includegraphics{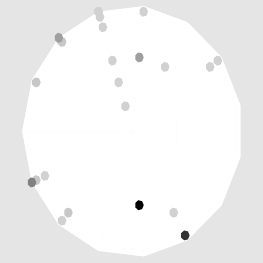}
		\includegraphics{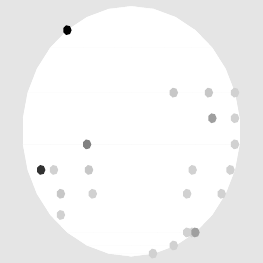}
 		\includegraphics{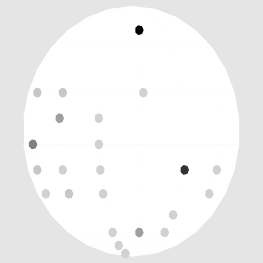}
	}

	\subfloat[][Estimation with proposals - Axial, coronal and sagittal views respectively]{
		\includegraphics{Figs/proposals_illustrating/inter_with_location_ax}
		\includegraphics{Figs/proposals_illustrating/inter_with_location_cor}
 		\includegraphics{Figs/proposals_illustrating/inter_with_location_sag}
	}	
	\caption{Illustration of inter-chain proposals effectiveness.}
	\label{fig:proposals_inter_prob_detection}
\end{figure}

The benefit of introducing inter-chain proposals is illustrated in Fig. \ref{fig:proposals_inter_prob_detection}. The first row of images of this figure displays the five non-zeros pressent in the ground truth. Without using inter-chain proposals the different chains arrive to different modes after 100.000 iterations as illustrated in the second row (which displays the probability of finding each dipole active with 8 parallel MCMC chains). The introduction of inter-chain proposals causes all the different chains to converge to the same (correct) value of $\zb$ in less than 5.000 iterations as illustrated in the third row.

\subsection{Estimators}
The point estimators used in this study are defined as follows 
 \begin{align}
	\centering
	\hat{\zb} & \triangleq \argmax_{\bar{\zb} \in \{0,1\}^N} \Big(\#\mathcal{M}({\bar{\zb}})\Big) \label{eq:z} \\	\hat{p} & \triangleq \frac{1} {\# \mathcal{M}(\hat{\zb})} \sum_{m \in \mathcal{M}(\hat{\zb})} p^{(m)} \label{eq:p}	
\end{align}
where $\mathcal{M}(\bar{\zb})$ is the set of iteration numbers $m$ for which the sampled variable $\zb^{(m)} = \bar{\zb}$ after the burn-in period and $p$ stands for any of the variables $\Xb$, $a$, $\sigma_n^2$, $\omega$ and $\taub^2$. Thus the estimator $\hat{\zb}$ \eqref{eq:z} corresponds to a maximum a posteriori estimator whereas the estimator used for all the other sampled variables \eqref{eq:p} is a minimum mean square error (MMSE) estimator.

However, it is interesting to note that the proposed method provides the full distribution of the unknown parameters and is not limited to point-estimate as the methods based on the $\ell_{21}$ mixed norm.

It will be shown in the next section that in certain conditions, such as low SNR, the a-posteriori distribution of $\zb$ has several values of comparable probability. These values of $\zb$ are usually minor variations of each other (changing one of the dipoles to a neighboring position for instance). In this case, after convergence the algorithm oscillates between several values of $\zb$ which allows the proposed method to identify several possible solutions (each of them corresponding to a different value of $\zb$) with their corresponding probabilities. This is an advantage over the mixed $\ell_{21}$ mixed norm method that is only able to provide a point-estimate of the solution.

\section{Experimental results}
\label{sec:exp_results}

\subsection{Synthetic data}
Synthetic data are first considered to compare the $\ell_{21}$ mixed norm approach with the proposed method using a 212-dipole Stok three-sphere head model \cite{stok1986inverse} with 41 electrodes. Three kinds of activations are performed: (1) three dipoles with low SNR, (2) five dipoles with high SNR and (3) multiple dipoles with high SNR.

\begin{figure}[]
	\centering
	\subfloat[][Ground truth]{
		\includegraphics{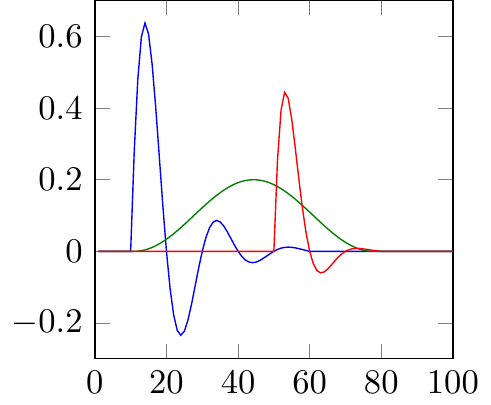}
	}
	\subfloat[][$\ell_{21}$-mixed norm estimation]{
		\includegraphics{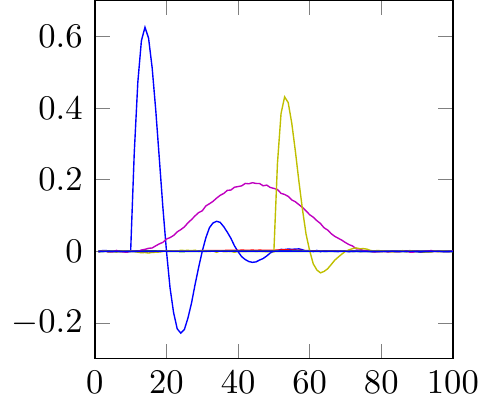}
	}
	\subfloat[][Proposed method]{
 		\includegraphics{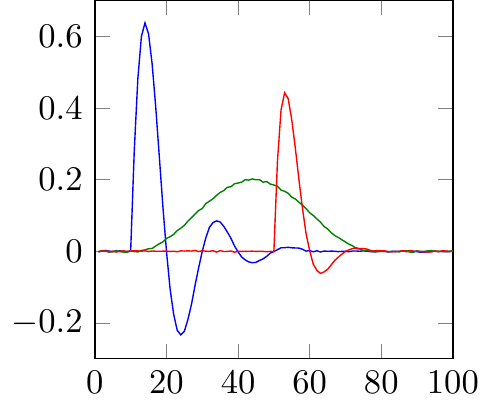}
	}
	
	\caption{Estimated waveforms for three dipoles with  SNR = 30dB.}
	\label{fig:three_dipoles_snr30_waveforms}
\end{figure}

\begin{figure}[]
	\centering
	\subfloat[][Ground truth - Axial, coronal and sagittal views respectively]{
		\includegraphics{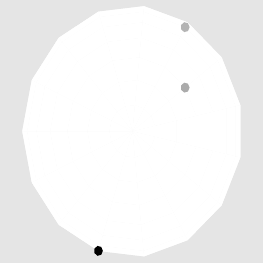}
		\includegraphics{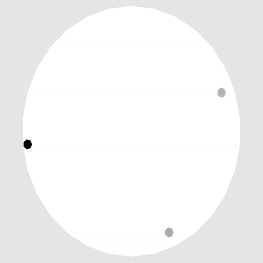}
 		\includegraphics{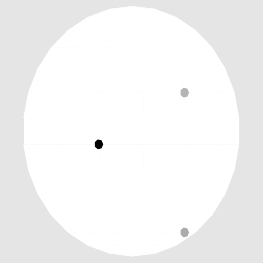}
	}

	\centering
	\subfloat[][$\ell_{21}$ - Axial, coronal and sagittal views respectively]{
		\includegraphics{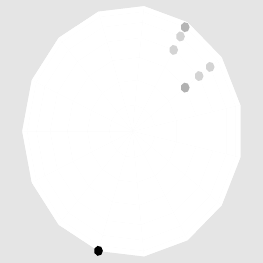}
		\includegraphics{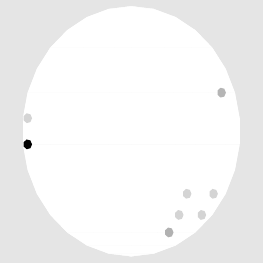}
 		\includegraphics{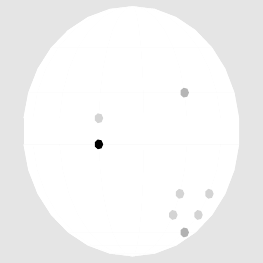}
	}
	
	\subfloat[][Proposed method - Axial, coronal and sagittal views respectively]{
		\includegraphics{Figs/three_dipoles/location/act_location_ax}
		\includegraphics{Figs/three_dipoles/location/act_location_cor}
 		\includegraphics{Figs/three_dipoles/location/act_location_sag}
	}
	
	\caption{Estimated activity for three dipoles and SNR = 30dB.}
	\label{fig:three_dipoles_snr30_locations}
\end{figure}

\begin{figure}[]
	\centering
	\subfloat[][Ground truth]{
		\includegraphics{Figs/three_dipoles/waveforms/act_waveform}
	}
	\subfloat[][$\ell_{21}$-mixed norm estimation]{
		\includegraphics{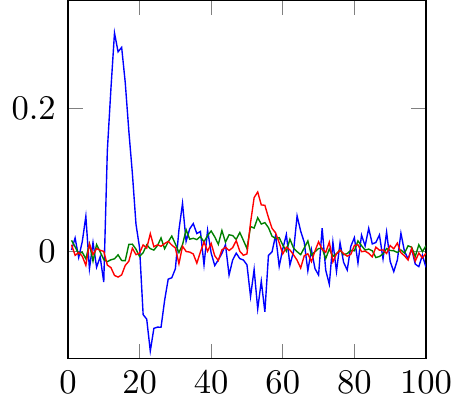}
	}
	\subfloat[][Proposed method]{
 		\includegraphics{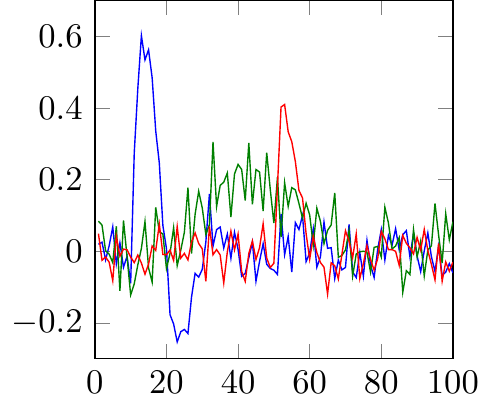}
	}
	
	\caption{Estimated waveforms for three dipoles with  SNR = -3dB.}
	\label{fig:three_dipoles_snrm3_waveforms}
\end{figure}

\begin{figure}[]
	\centering
	\subfloat[][Ground truth - Axial, coronal and sagittal views respectively]{
		\includegraphics{Figs/three_dipoles/location/act_location_ax}
		\includegraphics{Figs/three_dipoles/location/act_location_cor}
 		\includegraphics{Figs/three_dipoles/location/act_location_sag}
	}

	\subfloat[][$\ell_{21}$ - Axial, coronal and sagittal views respectively]{
		\includegraphics{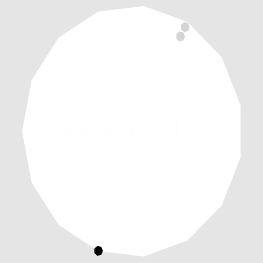}
		\includegraphics{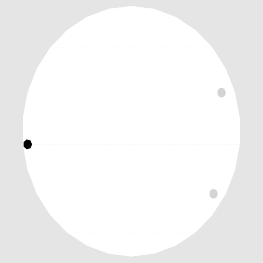}
		\includegraphics{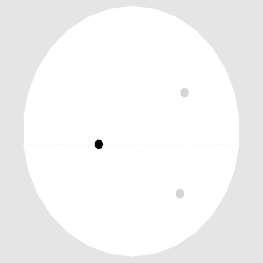}
	}
	
	\subfloat[][Proposed method - Axial, coronal and sagittal views respectively]{
		\includegraphics{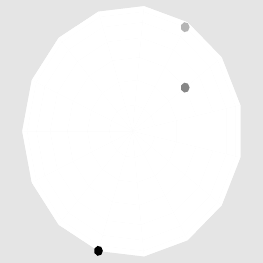}
		\includegraphics{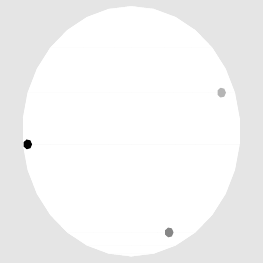}
 		\includegraphics{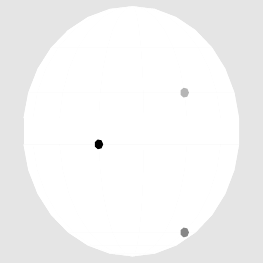}
	}
	
	\caption{Estimated activity for three dipoles and SNR = -3dB.}
	\label{fig:three_dipoles_snrm3_locations}
\end{figure}

\subsubsection{Three-dipoles with low SNR}
Three dipoles were assigned excitations defined as synthetic damped sinusoidal waves with frequencies between $5$ and $20$Hz. These excitations were $500$ms long (a period corresponding to a stationary dipole activity) and sampled at $200$Hz. Different levels of noise were used to compare the performance of the proposed method with the weighted $\ell_{21}$ mixed norm.  The parameters of our multiple dipole shift proposal were set to $K = 2$ and $\gamma = 0.8$ and eight MCMC chains were run in parallel. For the $\ell_{21}$ mixed norm approach, the value of the regularization parameter $\lambda$ was chosen using cross-validation to get the best possible result.

\begin{table}[]
	\centering
	\begin{tabular}{|c|c|}
	\hline\hline 
	Active non-zeros & Percentage of samples\\ [0.5ex]
	\hline
	1, 2, 3 & 43\%\\
	1, 2, 4 & 22\%\\
	1, 2, 5 & 11\%\\
	1, 2, 6 & 7\%\\
	1, 2, 7 & 6\%\\
	Others & 11\%\\
	\hline
	\end{tabular}
	\caption{Three dipoles with SNR = -$3$dB: modes explored after convergence. Positions 1, 2 and 3 correspond to the non-zero elements of the ground truth.}
	\label{fig:three_dipoles_snrm3_modes}
\end{table}

\begin{figure}[]
	\centering
	\subfloat[][Dipole waveform 1]{
		\includegraphics{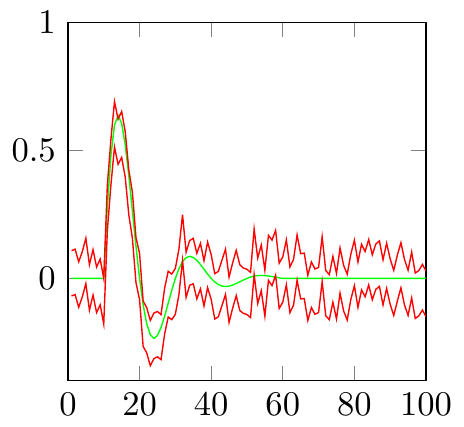}
	}
	\subfloat[][Dipole waveform 2]{
		\includegraphics{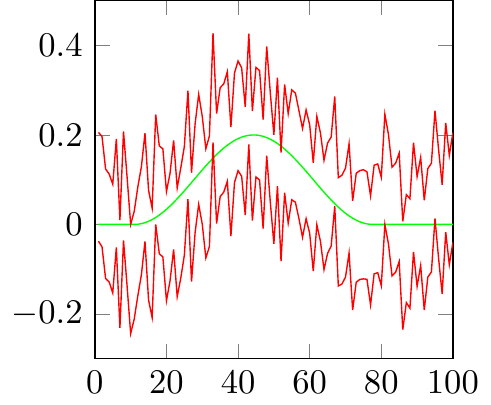}
	}
	\subfloat[][Dipole waveform 3]{
 		\includegraphics{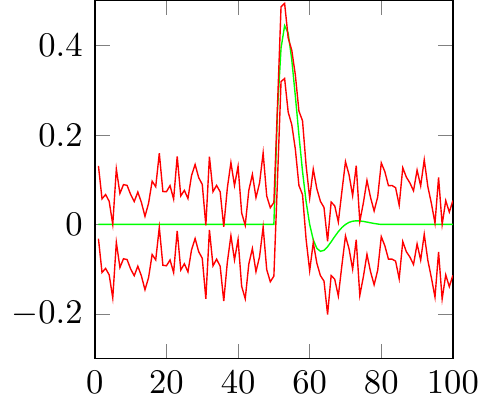}
	}
	
	\caption{Estimated boundaries $\mu \pm 2 \sigma$ for the three dipole simulation with SNR = -$3$dB.}
	\label{fig:three_dipoles_snrm3_waveforms_separate}
\end{figure}

The results for SNR = $30$dB are shown in Fig. \ref{fig:three_dipoles_snr30_waveforms} and Fig. \ref{fig:three_dipoles_snr30_locations}. Both algorithms seem to provide the same solution that is equal to the ground truth. The only minor difference is that the $\ell_{21}$-mixed norm regularization presents some dipoles (around ten) with very low but non-zero values, while our algorithm only detects the three non-zeros as real activity, this can be seen in the estimated dipoles locations in Fig. \ref{fig:three_dipoles_snr30_locations}.

The estimated dipole locations with SNR = $-3$dB are shown in Fig. \ref{fig:three_dipoles_snrm3_locations} whereas the corresponding estimated waveforms are shown in Fig. \ref{fig:three_dipoles_snrm3_waveforms}. Note that only the dipoles with highest activity are displayed for the $\ell_{21}$ approach. The approach based on the $\ell_{21}$ norm manages to recover only two of the three non-zero activities at the correct positions and seems to underestimate considerably the amplitude of the activity. This is a known problem caused by approximating the $\ell_0$ pseudo-norm by the $\ell_1$ norm, since the later penalizes high amplitudes while the former penalizes all non-zero values equally. Our algorithm oscillates between several values of $\zb$ (specified in Table \ref{fig:three_dipoles_snrm3_modes}). However, the most probable value of $\zb$ found by the algorithm is the correct one whereas the other ones have one of the active non-zeros moved to a close neighbor.

\begin{figure}[]
	\centering
	\subfloat[][Histogram of $\omega$]{
		\includegraphics{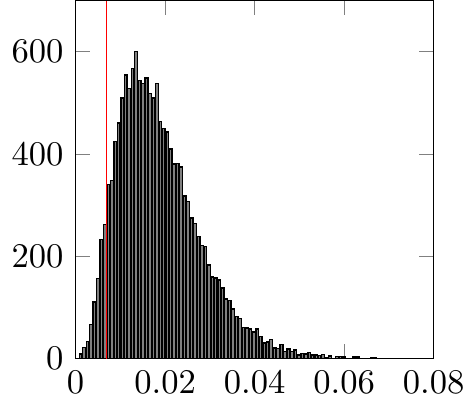}
	}
	\subfloat[][Histogram of $a$]{
		\includegraphics{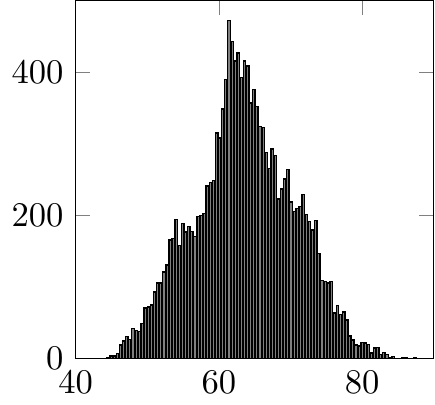}
	}
	\subfloat[][Histogram of $\sigma_n^2$]{
 		\includegraphics{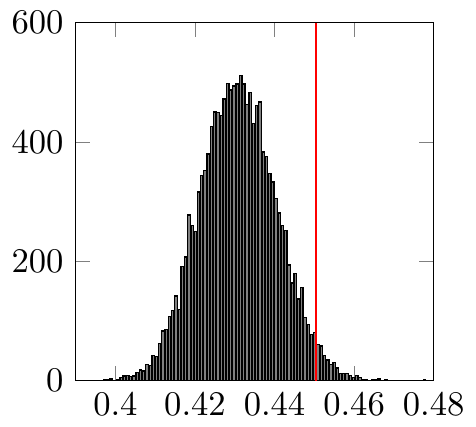}
	}
	
	\caption{Three dipoles with SNR = -$3$dB: histograms of the hyperparameters. The actual values of $\omega$ and $\sigma_n^2$ are marked with a red vertical line.}
	\label{fig:three_dipoles_snrm3_histograms}
\end{figure}

\begin{figure}[!]
	\centering
 	\subfloat[][PSRF of $\omega$]{
		\includegraphics{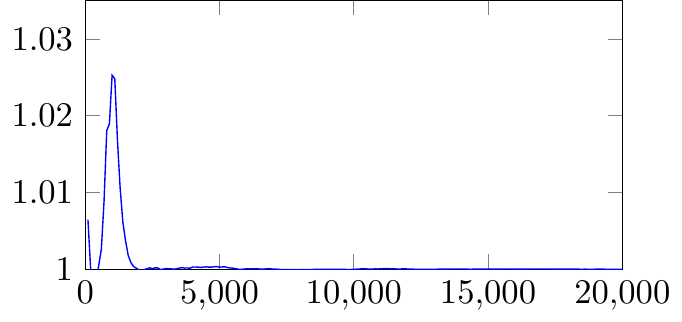}
	}
	\subfloat[][PSRF of $a$]{
		\includegraphics{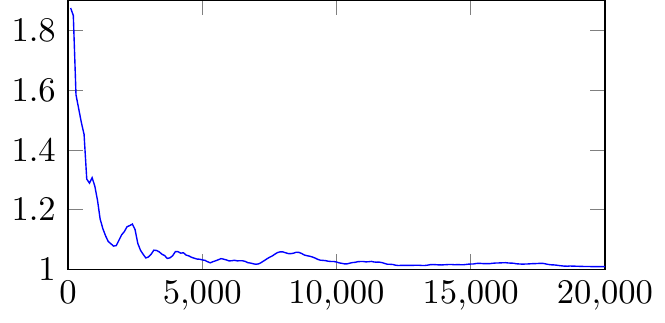}
	}
	
	\subfloat[][PSRF of $\sigma_n^2$]{
 		\includegraphics{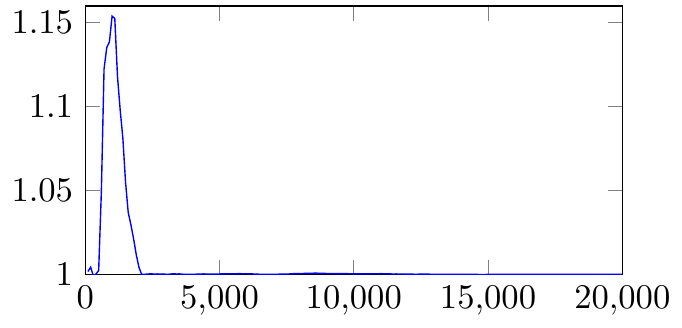}
	}
	\subfloat[][Maximum PSRF of $\Xb$]{
 		\includegraphics{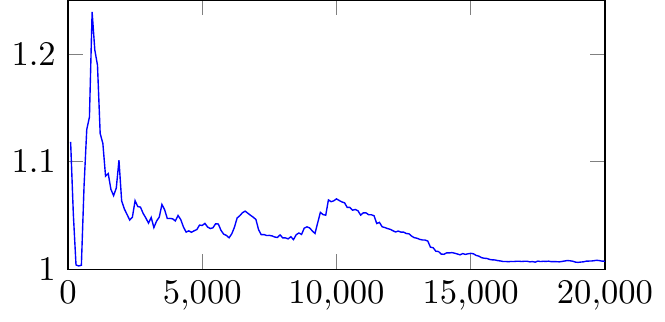}
	}
	\caption{Three dipoles with SNR = -$3$dB: PSRFs of sampled variables.}
	\label{fig:three_dipoles_snrm3_psrfs}
\end{figure}

The proposed method does not only allow a point-estimation of the activity but can also be used to estimate uncertainties associated with the activity. For instance, Fig. \ref{fig:three_dipoles_snrm3_waveforms_separate} shows the confidence intervals of the activity estimation (mean $\pm$ standard deviations). The actual ground truth activation is clearly located within two standard deviations of the estimated mean value obtained with the proposed algorithm. The histogram of the generated hyperparameters $\omega$, $a$ and $\sigma_n^2$ are shown in Fig. \ref{fig:three_dipoles_snrm3_histograms}. They are clearly in good agreement with the actual values of the corresponding parameters. The PSRF's are displayed in Fig. \ref{fig:three_dipoles_snrm3_psrfs}. It is possible to see that the PSRF's tend to 1 as the iterations increase, showing that the simulation is converging correctly.

\begin{figure}[]
	\centering
	\subfloat[][Ground truth - Axial, coronal and sagittal views respectively]{
		\includegraphics{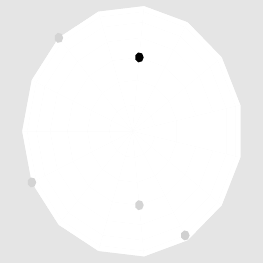}
		\includegraphics{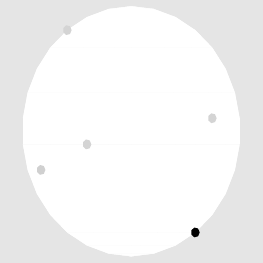}
 		\includegraphics{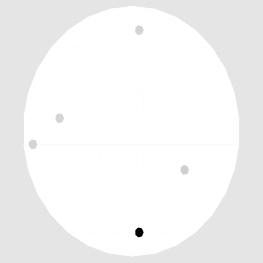}
	}

	\centering
	\subfloat[][$\ell_{21}$ dipole locations - Axial, coronal and sagittal views respectively]{
		\includegraphics{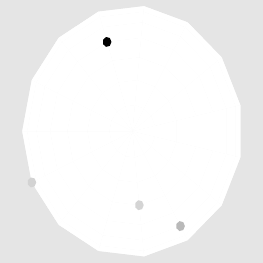}
		\includegraphics{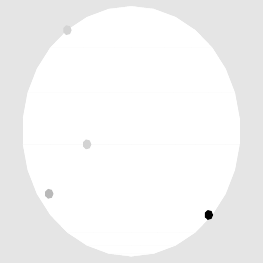}
 		\includegraphics{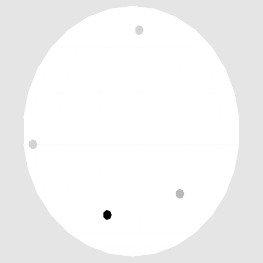}
	}

	\subfloat[][Proposed Method - Axial, coronal and sagittal views respectively]{
		\includegraphics{Figs/multiple_dipoles/location/l20_5_dipole_location_ax}
		\includegraphics{Figs/multiple_dipoles/location/l20_5_dipole_location_cor}
 		\includegraphics{Figs/multiple_dipoles/location/l20_5_dipole_location_sag}
	}
		
	\caption{Estimated activity for five dipoles and SNR = 30dB.}
	\label{fig:five_dipoles_locations}
\end{figure}

\begin{figure}[]
	\centering
	\subfloat[][Dipole waveform 1]{
		\includegraphics{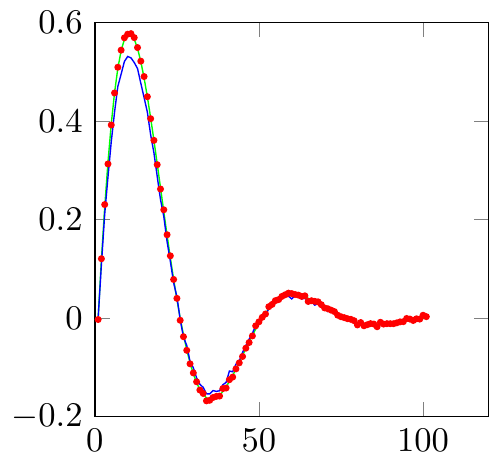}
	}
	\subfloat[][Dipole waveform 2]{
		\includegraphics{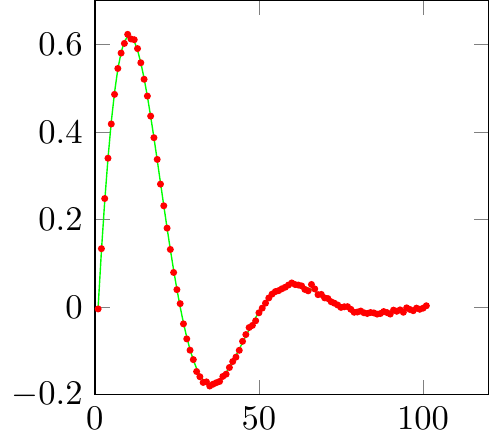}
	}
	\subfloat[][Dipole waveform 3]{
 		\includegraphics{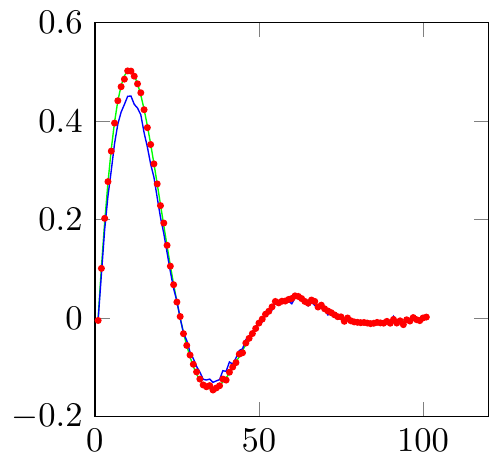}
	}

	\subfloat[][Dipole waveform 4]{
		\includegraphics{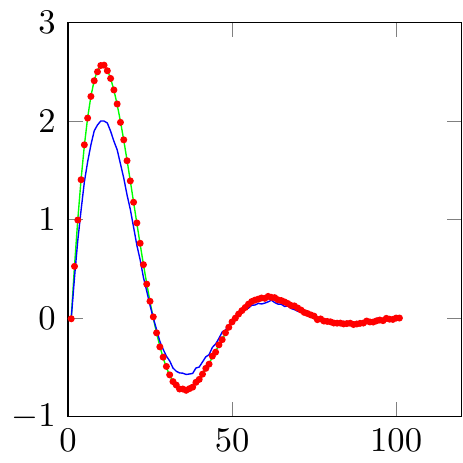}
	}
	\subfloat[][Dipole waveform 5]{
 		\includegraphics{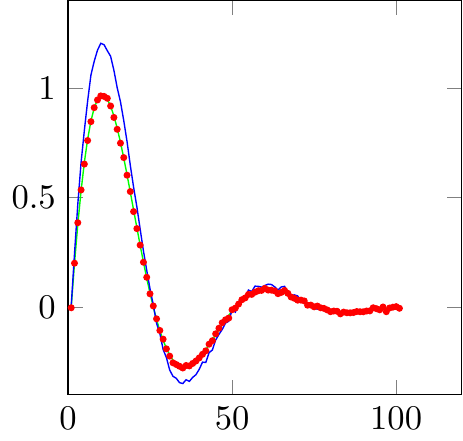}
	}

	\caption{Estimated waveforms for five dipoles with  SNR = 30dB. Green represents the ground truth, blue the $\ell_{21}$ mixed norm estimation and red the proposed method.}
	\label{fig:five_dipoles_waveforms}
\end{figure}

\subsubsection{Five dipoles}
On our second kind of experiments, five dipoles were activated with the same damped sinusoidal wave of 5Hz. The activations were sampled at 200Hz and scaled in amplitude so that each of them produced the same energy in the measurements. Noise was added to the measurements to obtain SNR = $30$dB. For the $\ell_{21}$ mixed norm regularization the regularization parameter was set according to the uncertainty principle which consists in finding a solution $\hat{\Xb}$ such that $||\Hb \hat{\Xb} - \Yb|| \approx ||\Hb \Xb - \Yb||$ \cite{morozov1966solution}. Eight MCMC chains were run in parallel for the proposed method. Only the five non-zeros of the estimated activity with highest energy in the measurements were considered.

The results are displayed on Fig. \ref{fig:five_dipoles_locations} and Fig. \ref{fig:five_dipoles_waveforms}. In the first one we are able to see that the proposed method is able to recover the five locations perfectly while the $\ell_{21}$ norm only detects four activations, two of which are not in the correct locations. In the waveforms displayed in Fig. \ref{fig:five_dipoles_waveforms} we can see that, while both methods are able to recover the general activity pattern, the proposed method closely matches the waveform amplitude while the $\ell_{21}$ mixed norm does not. This is partially due to the fact that it detects some of the activations in the wrong positions (waveforms 4 and 5) and partially due to the tendency to underestimate the activation amplitudes inherent to the $\ell_0$ to $\ell_1$ convex relaxation.

\begin{figure}[]
	\centering
 	\subfloat[][Recovery rate as a function of $P$]{
		\includegraphics{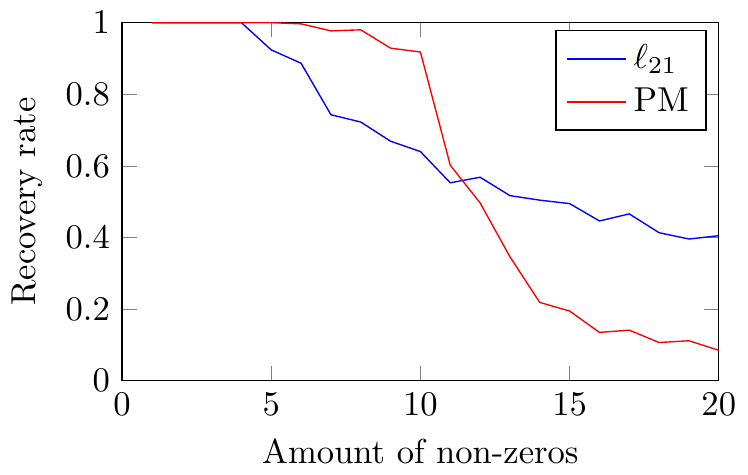}
	}
	\subfloat[][Proportion of residual energy as a function of $P$]{
		\includegraphics{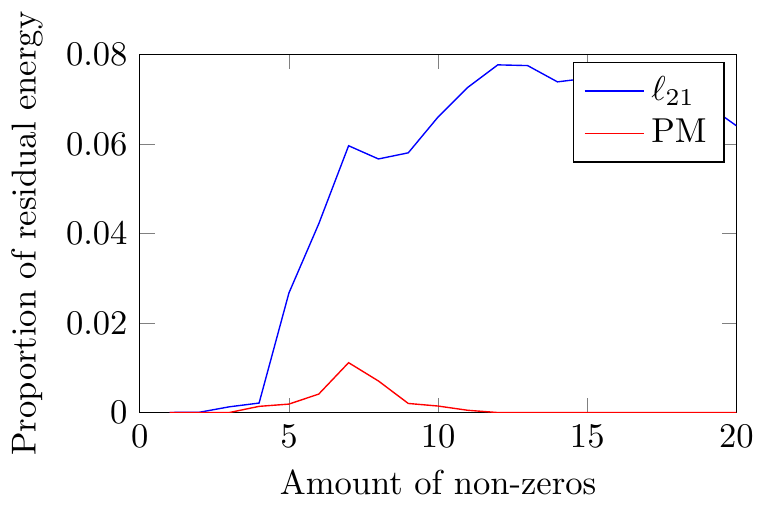}
	}
	\caption{Performance measures for multiple dipoles.}
	\label{fig:multiple_dipoles_rec_rate_residual_energy}
\end{figure}

\subsubsection{Multiple dipoles}
In this section, we compare the detection capabilities of the algorithm with respect to the $\ell_{21}$-mixed norm approach by varying the amount of non-zeros present in the ground truth.

In each simulation of this section, $P$ dipoles were activated with damped sinusoidal waves with frequencies varying between 5 and 20Hz. The activations were sampled at 200Hz and scaled in amplitude so that each of them produced the same energy in the measurements. Fifty different sets of localizations were used for the non-zero positions for each value of $P=1,...,20$, resulting in a total of $1000$ experiments. Noise was added to the measurements to obtain SNR = $30$dB. For the $\ell_{21}$ mixed norm regularization the regularization parameter was set according to the uncertainty principle.

For each simulation, the $P$ non-zeros of the estimated activity associated with the highest energy in the measurements were considered as the estimated activity whereas the other elements were considered as residuals. We define the recovery rate as the proportion of non-zeros in the ground truth that are also present in the estimated activity. The average recovery rates of the proposed method and the $\ell_{21}$ mixed norm approach are presented in the first plot of Fig. \ref{fig:multiple_dipoles_rec_rate_residual_energy} as a function of $P$.
For $P \leq 10$ our algorithm detects the non-zeros with an accuracy higher than 90\% which drops to 60.2\% for $P = 11$ and 49.7\% for $P = 12$. This  drop of the recovery rate when a large number of non-zeros is present in the ground truth is well known, since the possible amount of non-zeros to recover correctly is limited by the span of the operator \cite{candes2008restricted}. For comparison, the $\ell_{21}$ mixed norm regularization recovers up to $P = 5$ non-zeros with an accuracy higher than 90\% and its recovery rate decreases slowly to reach 64\% for $P = 10$. Note that our method performs better than the $\ell_{21}$ approach for $P \leq 11$. However, beyond this point, the performance of both methods is very poor preventing them from being used in practical applications.

The recovery rate is calculated from the $P$ main non-zero elements of the activity. However, it is also interesting to analyze how much activity is present in the residual non-zero elements. Thus, we define the proportion of residual energy as the amount of energy contained in the measurements generated by the residual non-zeros with respect to the total energy in the measurements. This residual energy serves as a measure of the sparsity of the solution.  The second plot of Fig. \ref{fig:multiple_dipoles_rec_rate_residual_energy} shows the value  of the residual energy obtained for both algorithms as a function of $P$. The $\ell_{21}$ approach has up to 7.7\% of the activity detected in residual non-zeros whereas our algorithm never exceeds 1.1\% and always has lower residual activity than $\ell_{21}$, confirming its good sparsity properties.

\subsection{Real data experiment}
Two real data sets are then considered to validate the proposed method. The first dataset corresponds to the auditory evoked responses to left ear pure tone stimulus while the second one consists of the evoked responses to facial stimulus. The results of the proposed method are compared with the weighted $\ell_{21}$ mixed norm \cite{gramfort2012mixed} and the multiple sparse priors method \cite{friston2008multiple}.

\begin{figure}[]
	\centering
	\subfloat[][EEG whitened measurements]{
		\includegraphics{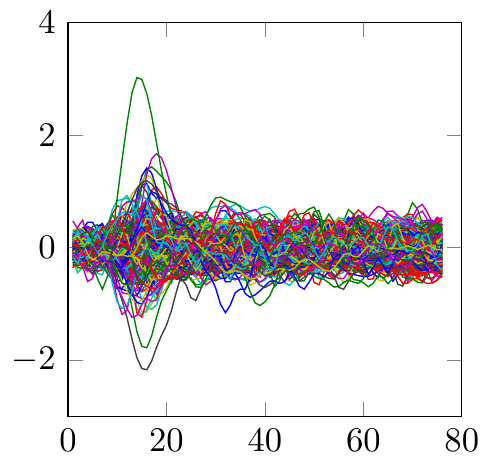}
	}
	\subfloat[][$\ell_{21}$-mixed norm estimated waveforms]{
		\includegraphics{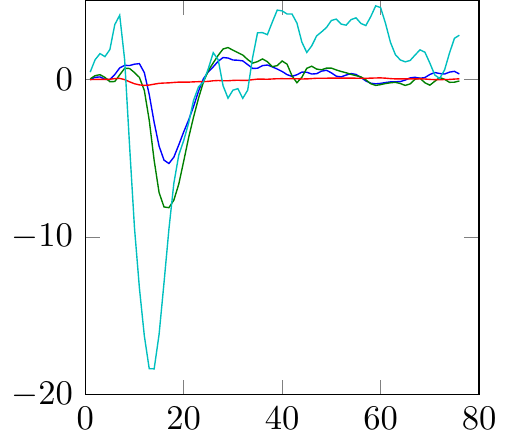}
	}
	\subfloat[][Proposed method estimated waveforms]{
 		\includegraphics{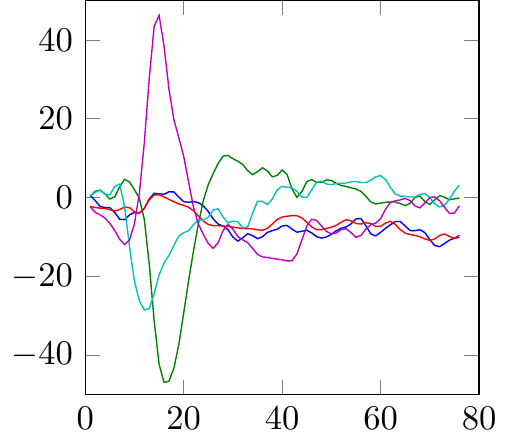}
	}
	
	\caption{Measurements and estimated waveforms for the auditory evoked responses.}
	\label{fig:real_data_waveforms_all}
\end{figure}

\begin{figure}[]
	\centering
	\subfloat[][Weighted $\ell_{21}$ mixed norm - Uncertainty principle of parameter]{
		\includegraphics[width=100pt, height=100pt]{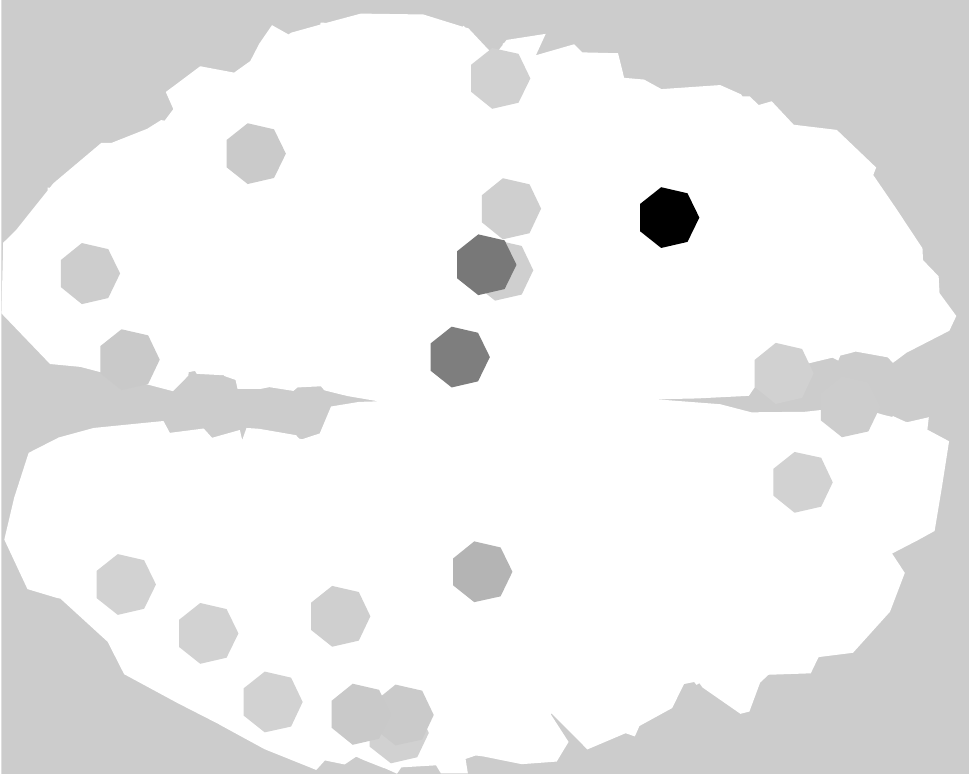}
		\includegraphics[width=100pt, height=100pt]{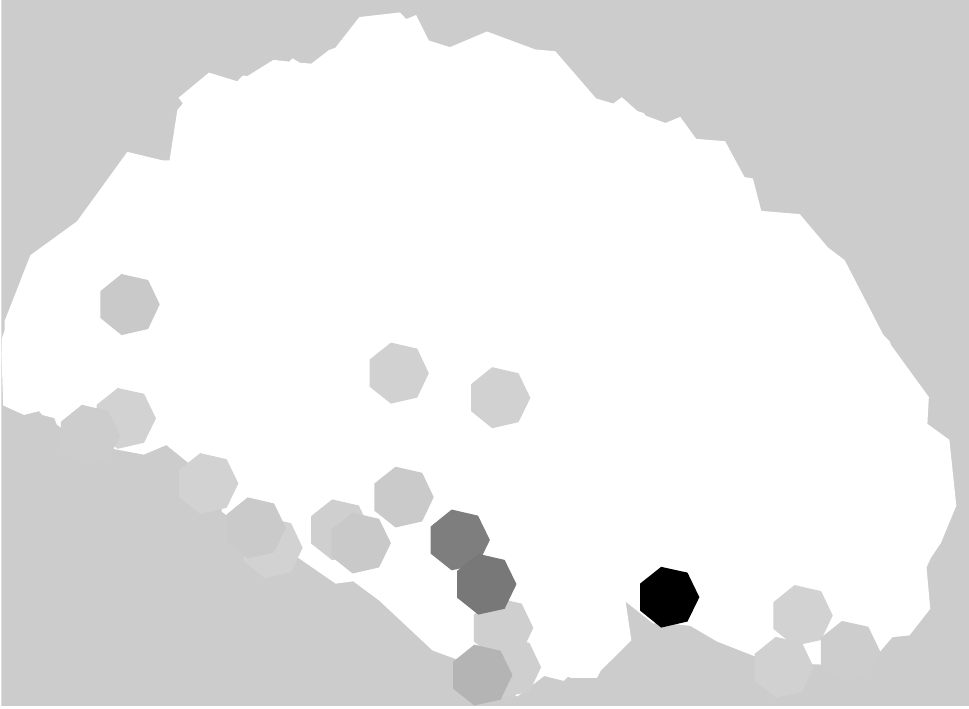}
		\includegraphics[width=100pt, height=100pt]{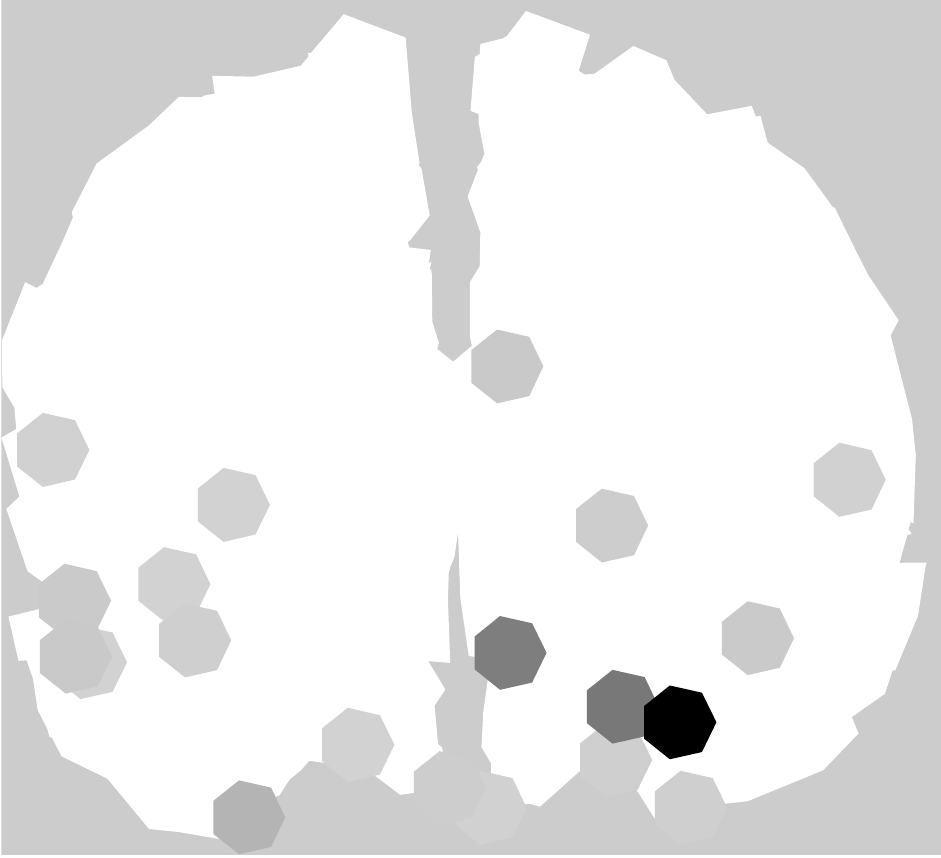}
	}
	
	\subfloat[][Weighted $\ell_{21}$ mixed norm - Manual adjustment of parameter]{
		\includegraphics[width=100pt, height=100pt]{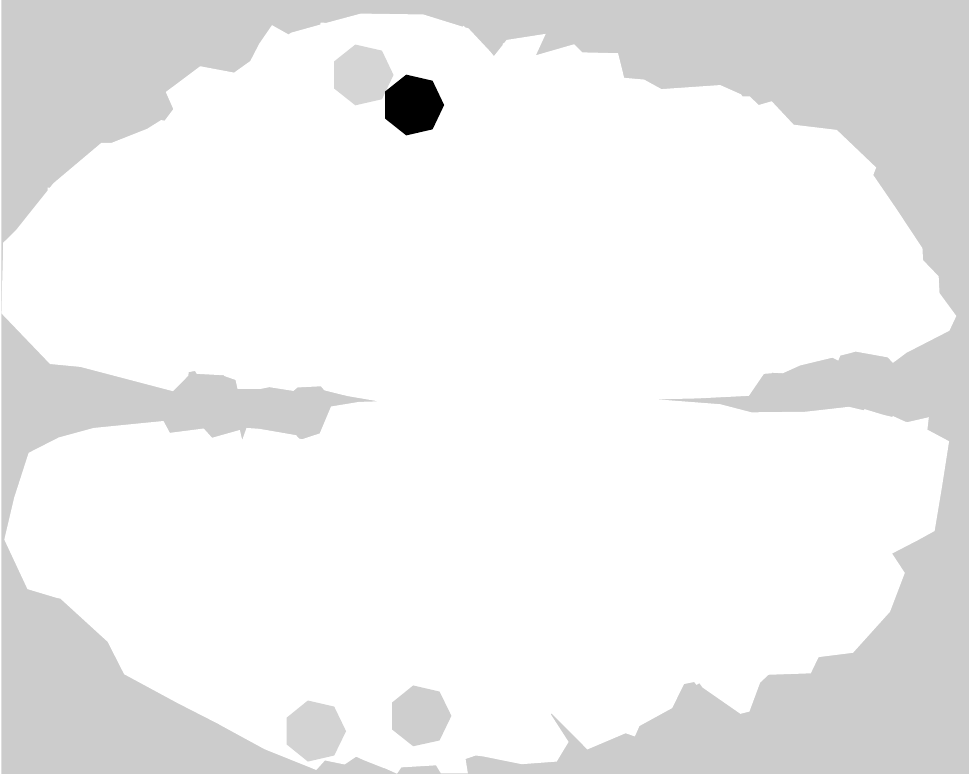}
		\includegraphics[width=100pt, height=100pt]{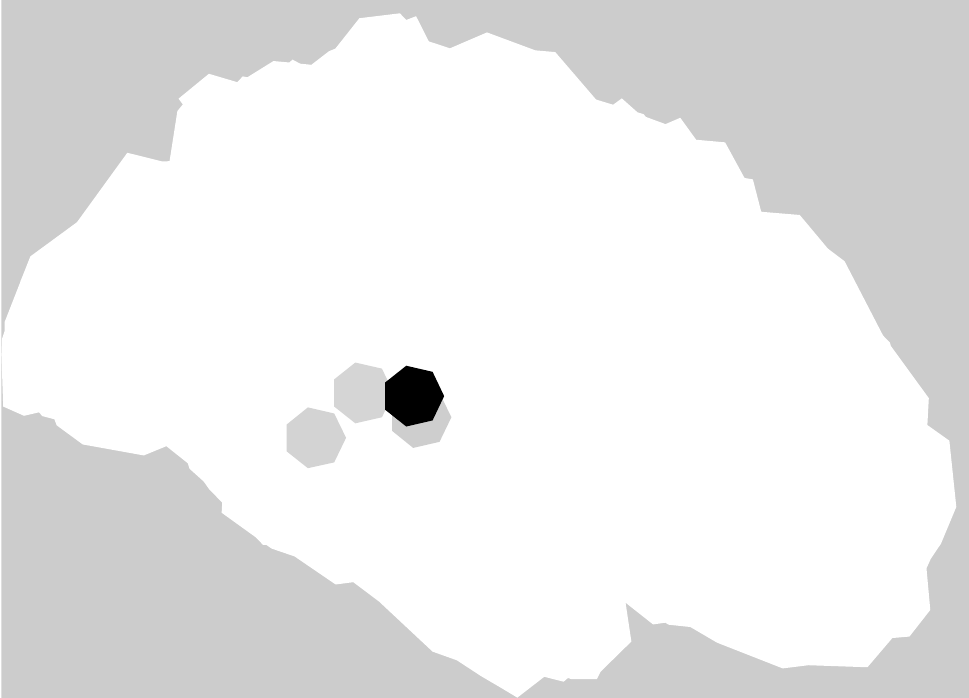}
		\includegraphics[width=100pt, height=100pt]{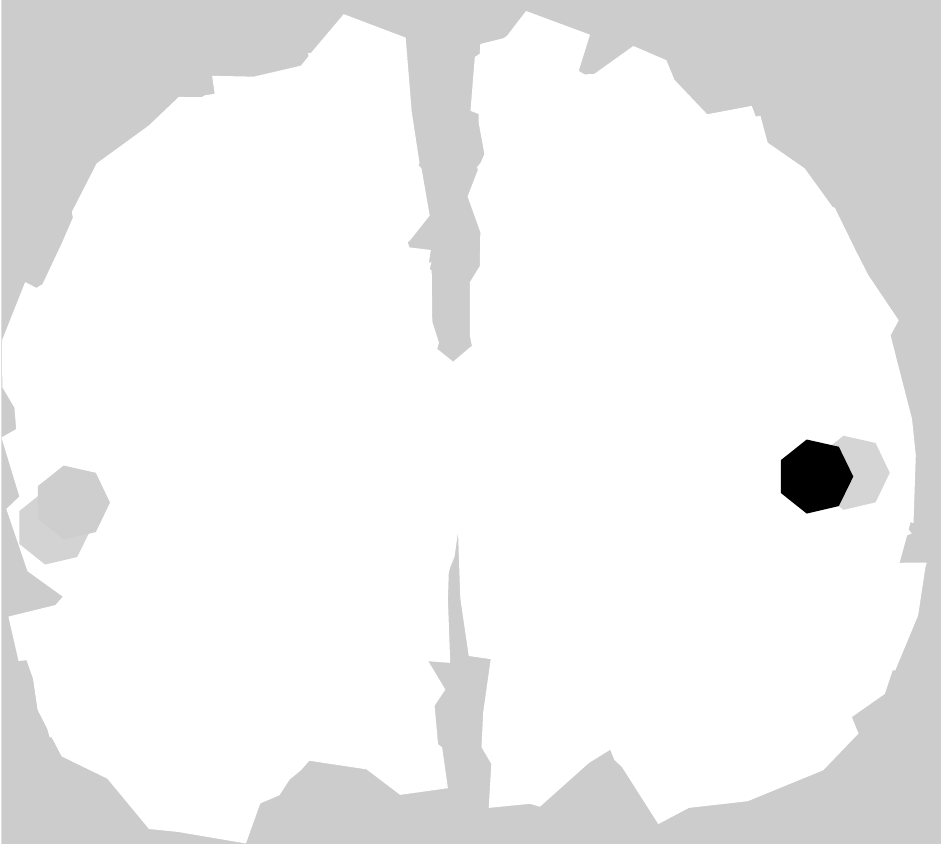}
	}
	
	\centering
	\subfloat[][Proposed method]{
		\includegraphics[width=100pt, height=100pt]{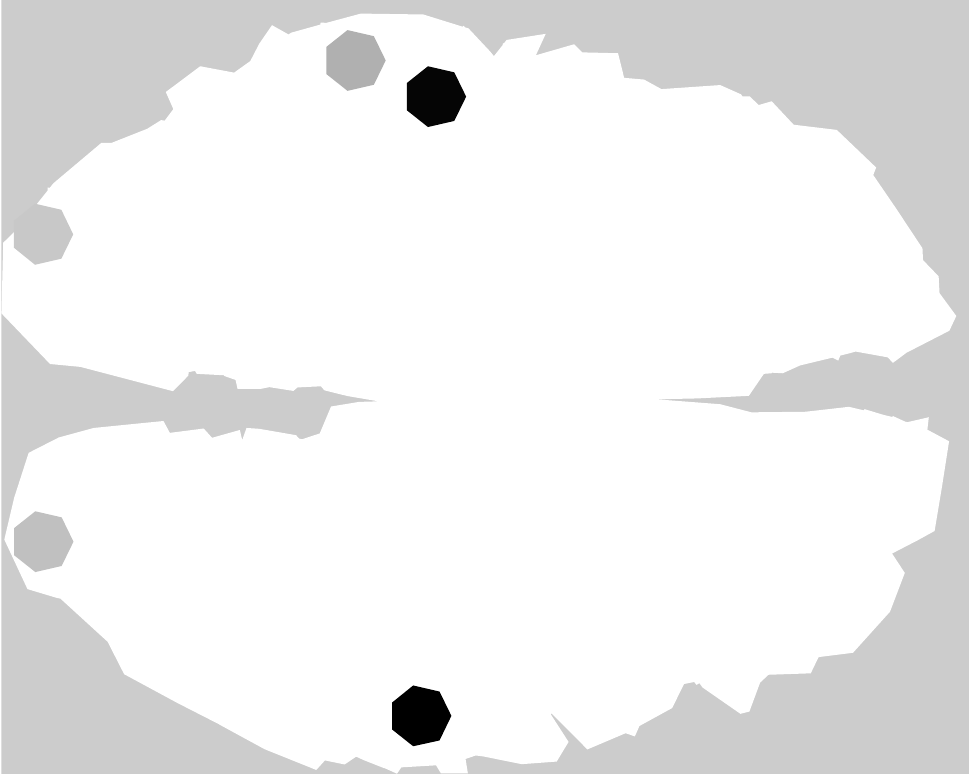}
		\includegraphics[width=100pt, height=100pt]{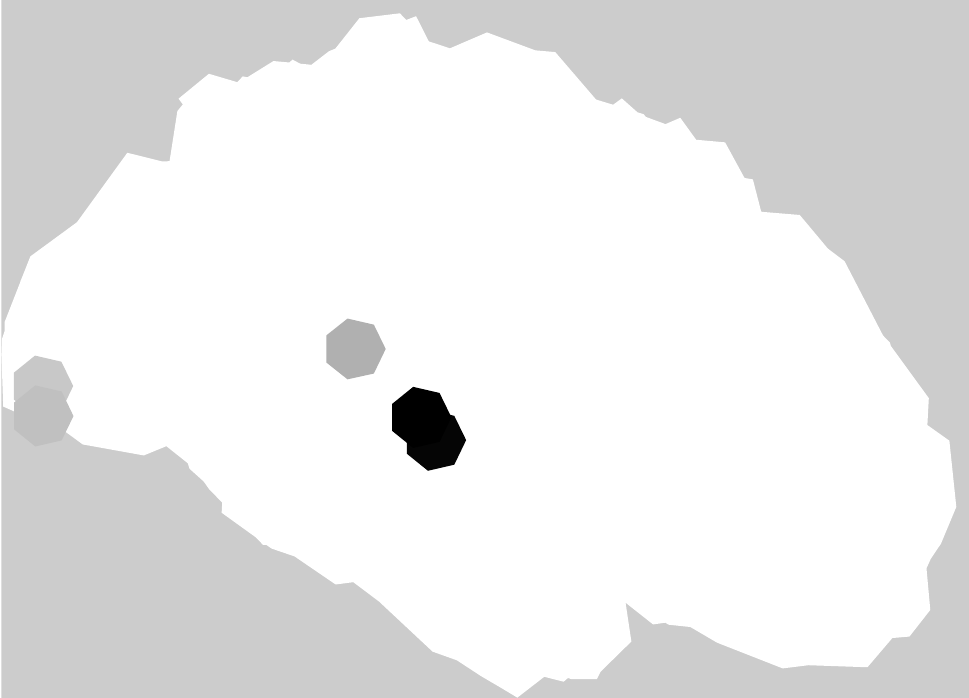}
		\includegraphics[width=100pt, height=100pt]{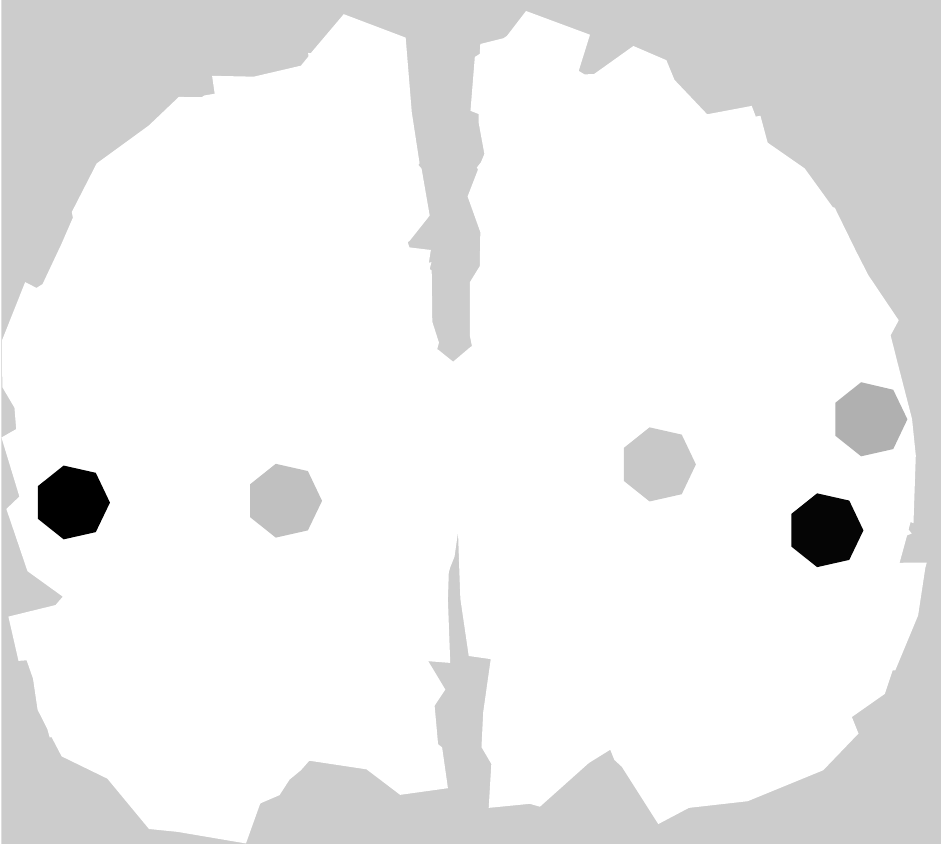}
	}

	\centering
	\subfloat[][MSP algorithm]{
		\includegraphics[width=100pt, height=100pt]{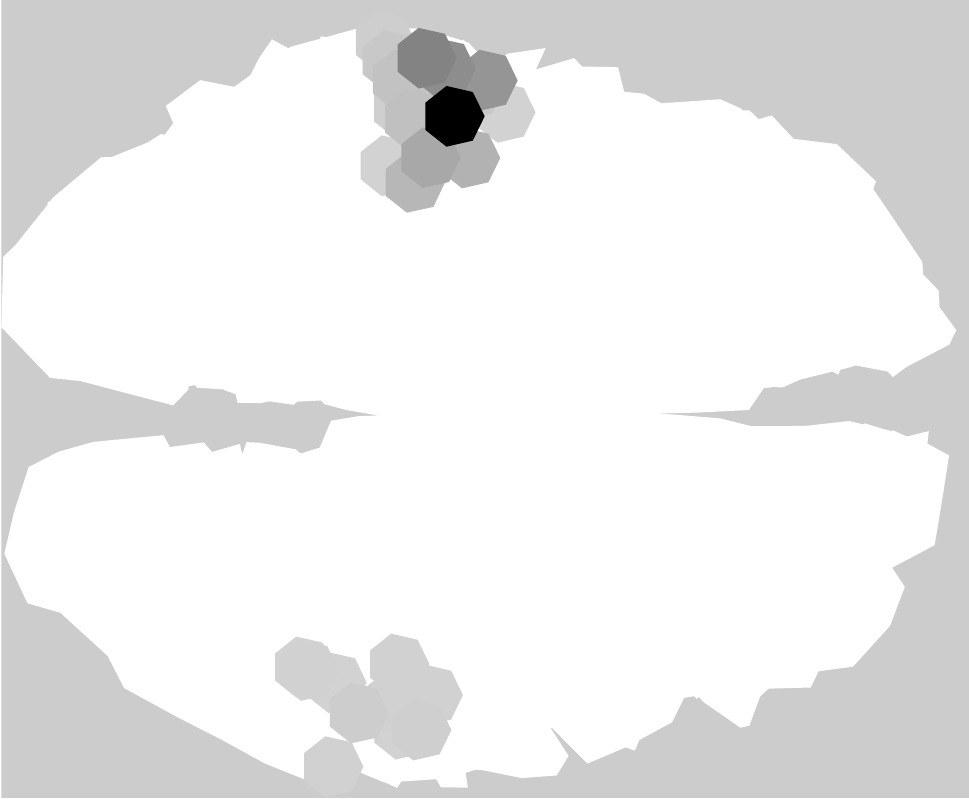}	
		\includegraphics[width=100pt, height=100pt]{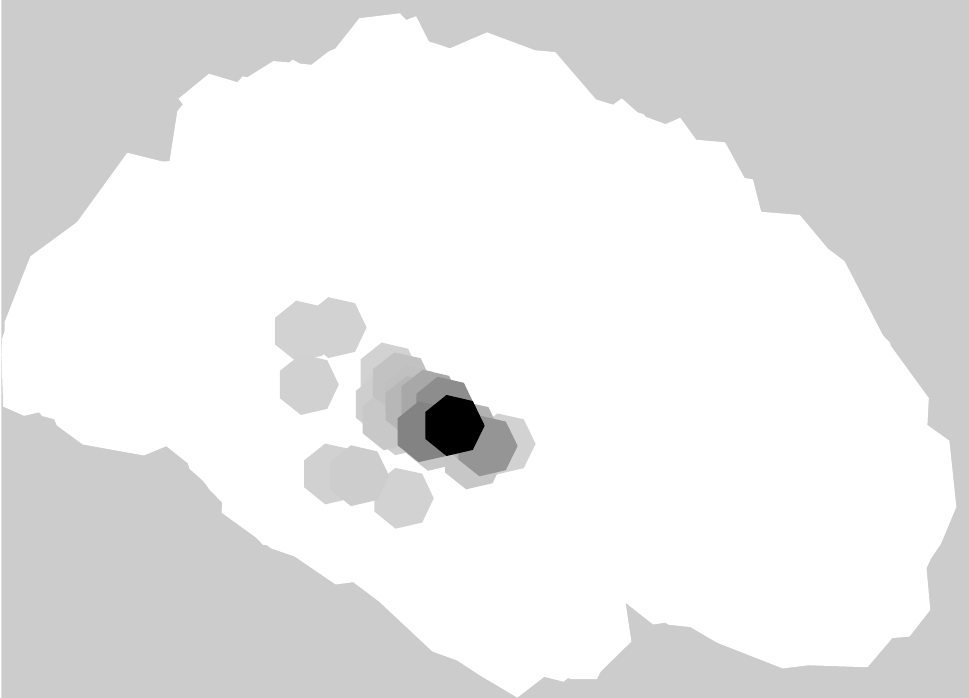}
		\includegraphics[width=100pt, height=100pt]{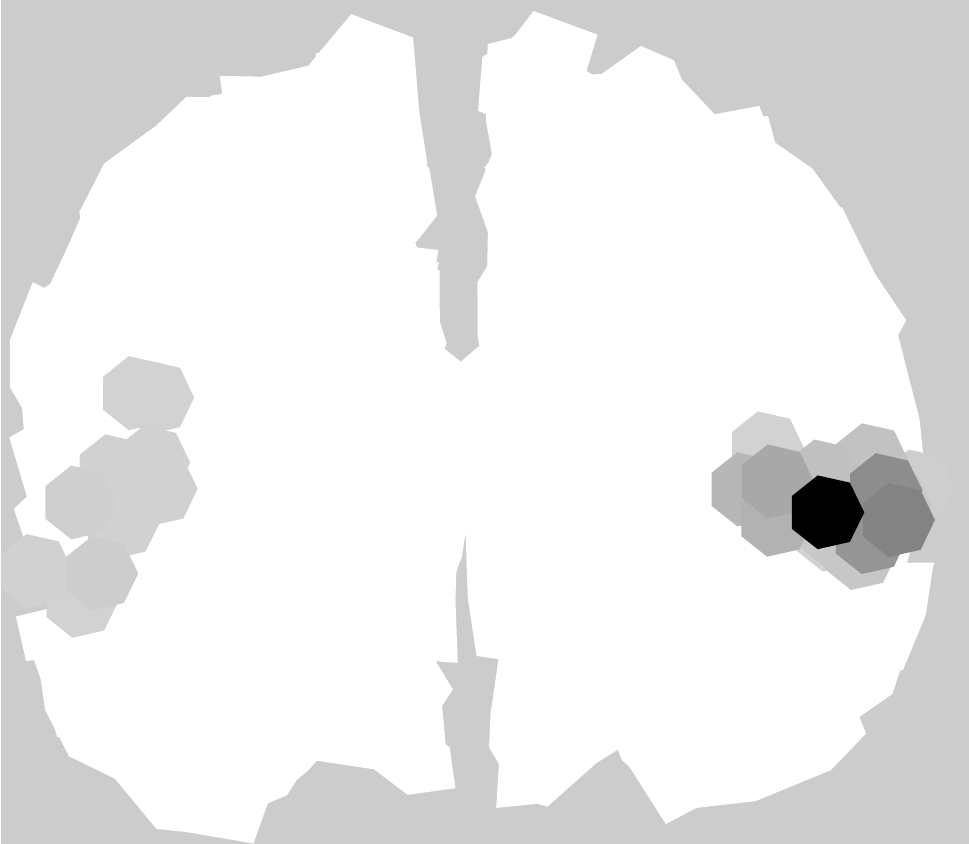}
	}
	
	\caption{Estimated activity for the auditory evoked responses.}
	\label{fig:real_data_location}
\end{figure}

\begin{figure}[]
	\centering
	\subfloat[][Waveform 1 (Center right dipole)]{
		\includegraphics{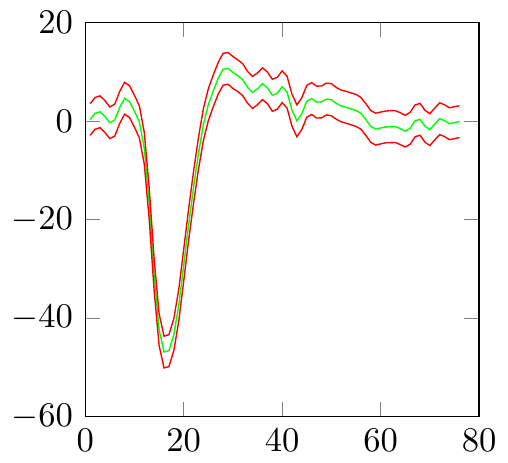}
	}
	\subfloat[][Waveform 2 (Center left dipole 1, closer to the edge)]{
		\includegraphics{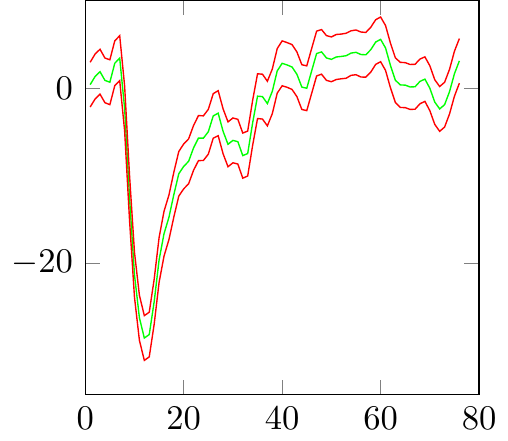}
	}
	\subfloat[][Waveform 3 (Center left dipole 2)]{
 		\includegraphics{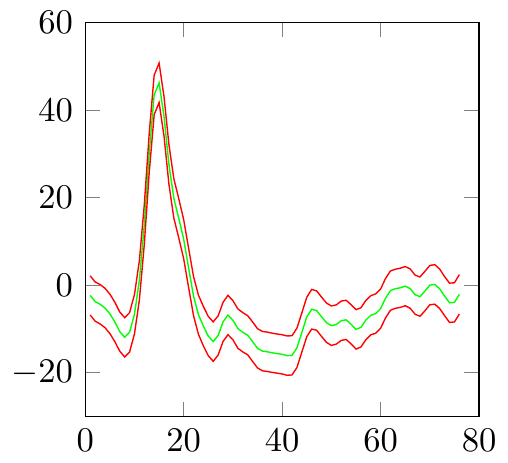}
	}		
	
	\subfloat[][Waveform 4 (Posterior right dipole)]{
		\includegraphics{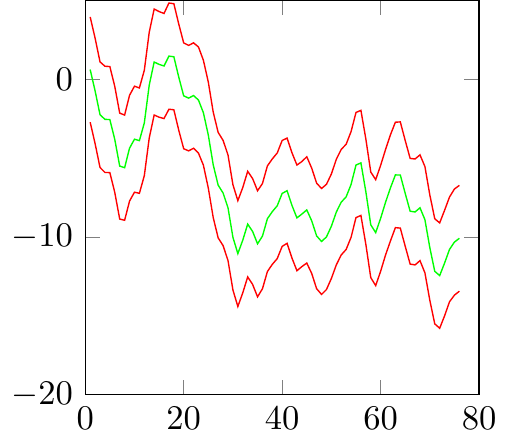}
	}	
	\subfloat[][Waveform 5 (Posterior left dipole)]{
 		\includegraphics{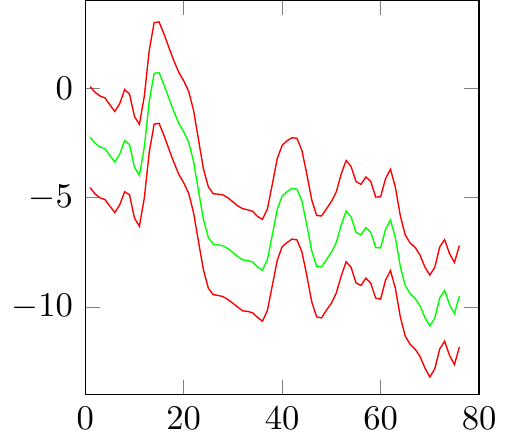}
	}
	
	\caption{Estimated waveforms mean and boundaries $\mu \pm 2 \sigma$ for the auditory evoked responses.}
	\label{fig:real_data_waveforms_separate}
\end{figure}

\begin{figure}[!]
	\centering
	\subfloat[][Histogram of $\omega$]{
		\includegraphics{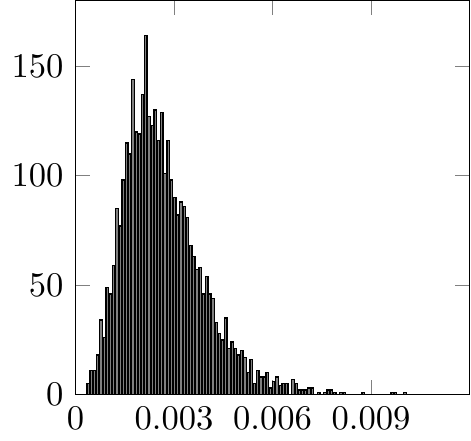}
	}
	\subfloat[][Histogram of $a$]{
		\includegraphics{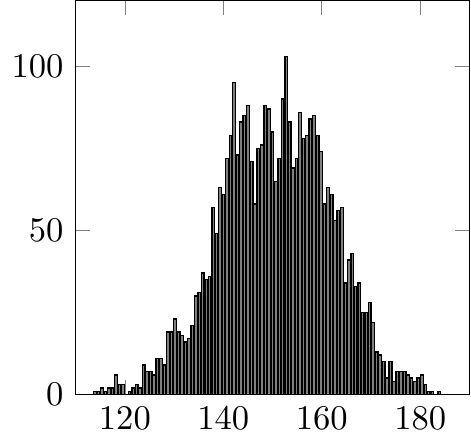}
	}
	\subfloat[][Histogram of $\sigma_n^2$]{
 		\includegraphics{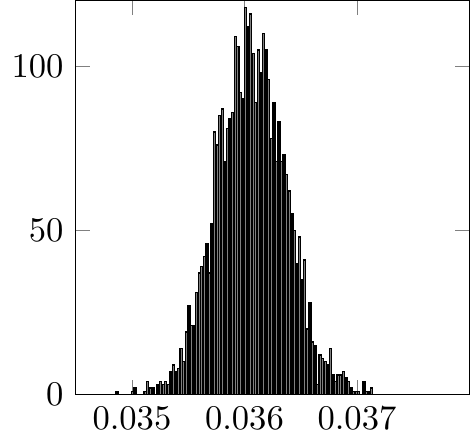}
	}
	
	\caption{Hyperparameters histograms for the auditory evoked responses.}
	\label{fig:real_data_histograms}
\end{figure}

\subsubsection{Auditory evoked responses}
The default data set of the MNE software \cite{gramfort2014mne,gramfort2013meg} is used in this section. It consists of the evoked response to left-ear auditory pure-tone stimulus using a realistic BEM (Boundary element method) head model sampled with $60$ EEG electrodes and $306$ MEG sensors. The  head model contains $1.844$ dipoles located on the cortex with orientations that are normal to the brain surface. Two channels that had technical artifacts were ignored. The data was sampled at $600$Hz. The samples were low-pass filtered at $40$Hz and downsampled to $150$Hz. The noise covariance matrix was estimated from $200$ms of the data preceding each stimulus and was used to whiten the measurements. Fifty-one epochs were averaged to calculate the measurements $\Yb$. The activity of the source dipoles was estimated jointly for the period from $0$ms to $500$ms after the stimulus. From a clinical perspective it is expected to find the brain activity primarily focused on the auditory cortices that are located close to the ears in both hemispheres of the brain.

\begin{figure}[!]
	\centering
	\subfloat[][PSRF of $\omega$]{
		\includegraphics{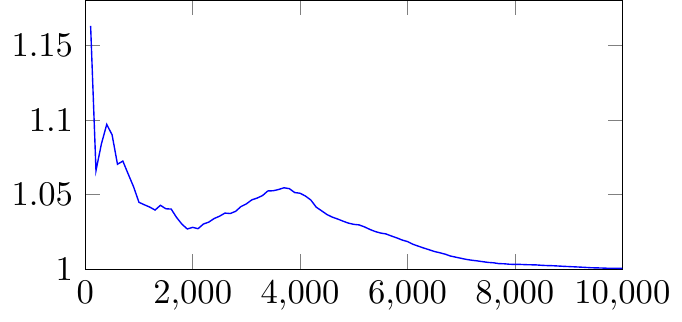}
	}
	\subfloat[][PSRF of $a$]{
		\includegraphics{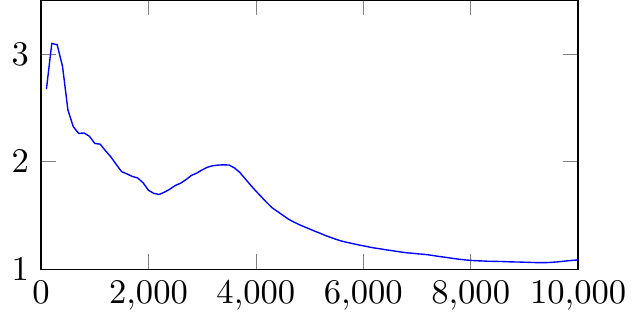}
	}

	\subfloat[][PSRF of $\sigma_n^2$]{
 		\includegraphics{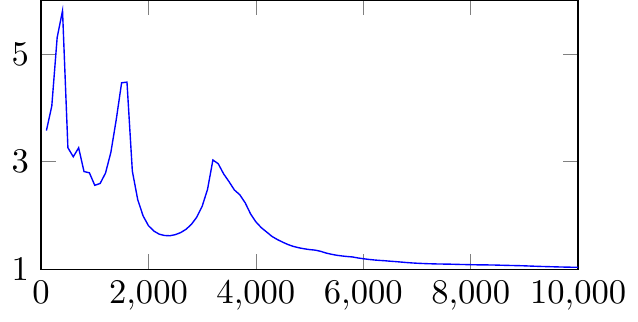}
	}
	\subfloat[][Maximum PSRF of $\Xb$]{
 		\includegraphics{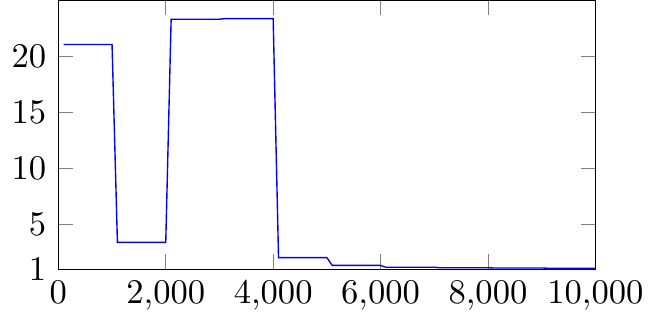}
	}
		
	\caption{PSRFs of sampled variables for the auditory evoked responses .}
	\label{fig:real_data_psrfs}
	
\end{figure}

Since the measurements were whitened, it is possible to use the uncertainty principle to adjust the hyperparameter of the $\ell_{21}$ mixed norm. However, this provides an activity distributed all over the brain as shown in the first row of Fig. \ref{fig:real_data_location}. By manually adjusting the hyperparameter to produce a sparser result, the $\ell_{21}$ mixed norm can obtain a solution that has activity in the auditory cortices as expected, shown in the second row of images. In contrast, our algorithm estimates its hyperparameters automatically and finds most of the activity in the auditory cortices without requiring any manual adjustment as displayed in the third row. On the other hand, the MSP method spreads the activity around the auditory cortices area since it groups the dipoles together in pre-defined regions.

The whitened measurements are displayed in Fig. \ref{fig:real_data_waveforms_all} along with the activity estimation for both the $\ell_{21}$ approach (after manually adjusting the hyperparameter) and our algorithm.

The five waveforms estimated by the proposed method with their confidence intervals of $2 \sigma$ are shown in Fig. \ref{fig:real_data_waveforms_separate}. Both results present sharp peaks in the activations of the auditory cortex dipoles between $80$ and $100$ milliseconds after the application of the stimulus. Note that the amplitudes estimated by the proposed method are much higher than the ones obtained with the $\ell_{21}$ approach due to the aforementioned amplitude underestimation of the latter.

The histograms of the hyperparameters of our algorithm are presented in Fig. \ref{fig:real_data_histograms} while their PSRF's are shown in Fig. \ref{fig:real_data_psrfs}. In the PSRF's we can see very abrupt changes in values for all the variables (most noticiably for $\Xb$) in the same iterations. These correspond to the iterations in which proposals were accepted by different chains and reflect the fact that the PSRF's can have very high values while the chains are in different modes. However, at the end of the simulation all chains converge to the same value of $\zb$ which causes all the PSRF's to tend to 1, showing the correct convergence of all the chains to the same posterior distribution.

\begin{figure}[]
	\centering
	\subfloat[][EEG whitened measurements]{
		\includegraphics{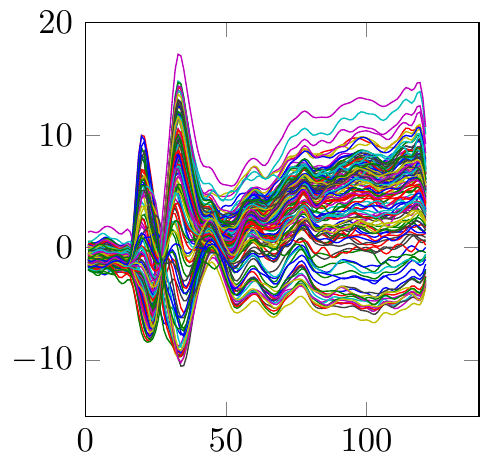}
	}
	\subfloat[][$\ell_{21}$-mixed norm estimated waveforms]{
		\includegraphics{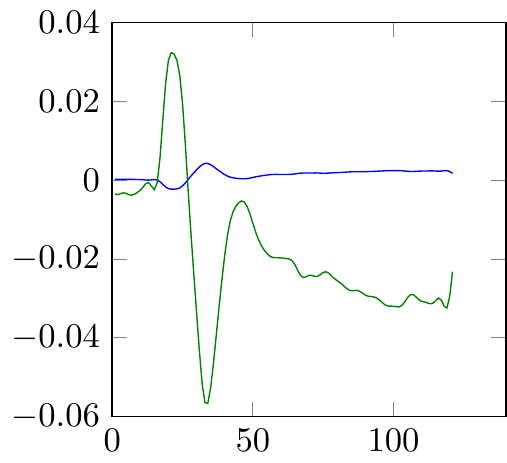}
	}
	\subfloat[][Proposed method estimated waveforms]{
 		\includegraphics{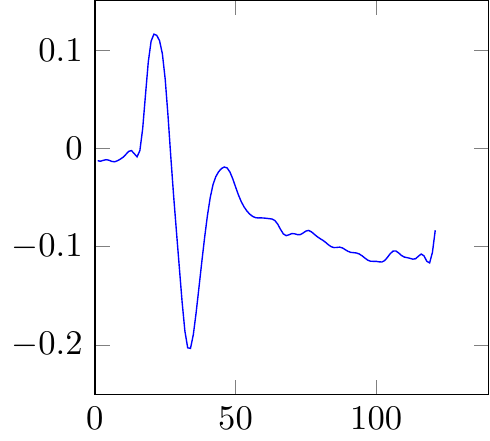}
	}
	
	\caption{Measurements and estimated waveforms for the facial evoked responses.}
	\label{fig:real_data_waveforms_all_facial}
\end{figure}

\subsubsection{Facial evoked responses}
\begin{figure}[]
	\centering
	\subfloat[][Weighted $\ell_{21}$ mixed norm - Uncertainty principle for parameter]{
		\includegraphics[width=100pt, height=100pt]{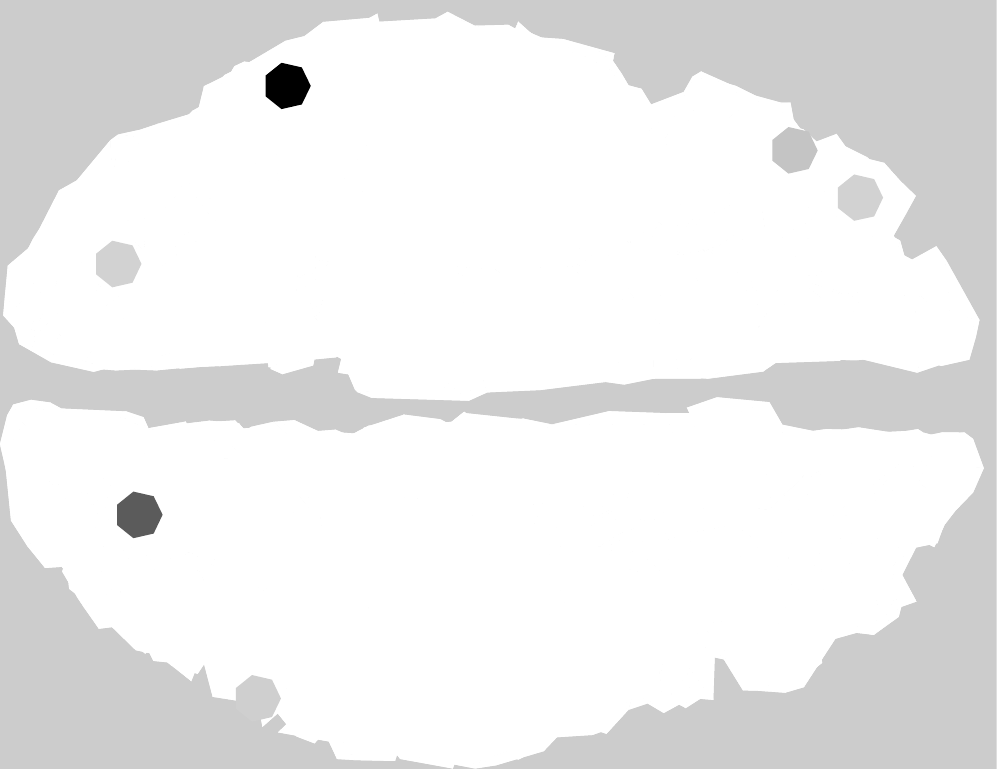}
		\includegraphics[width=100pt, height=100pt]{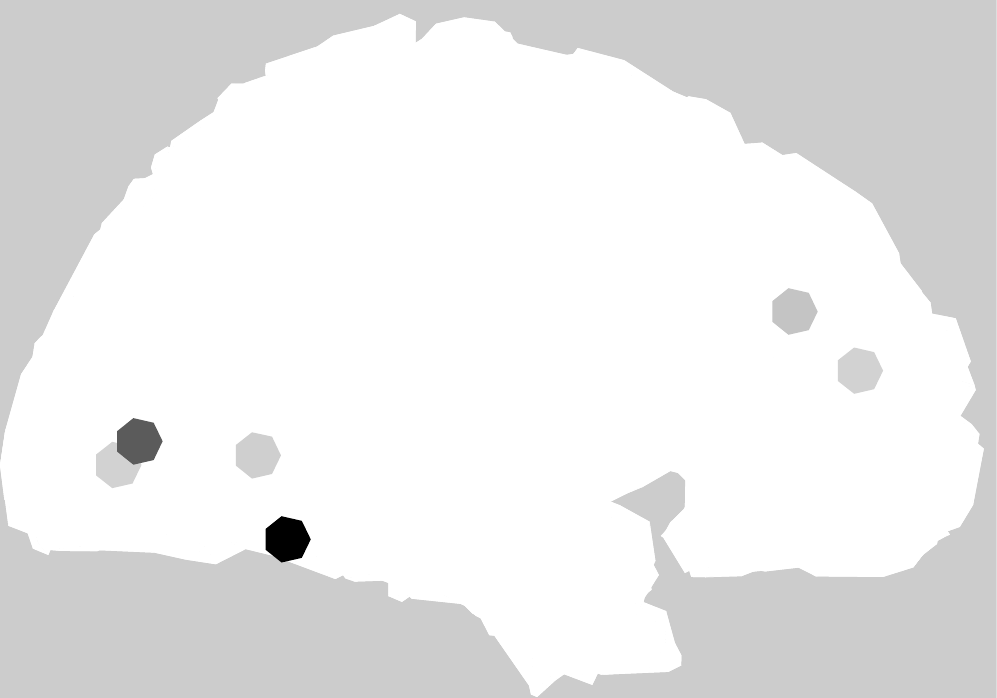}
		\includegraphics[width=100pt, height=100pt]{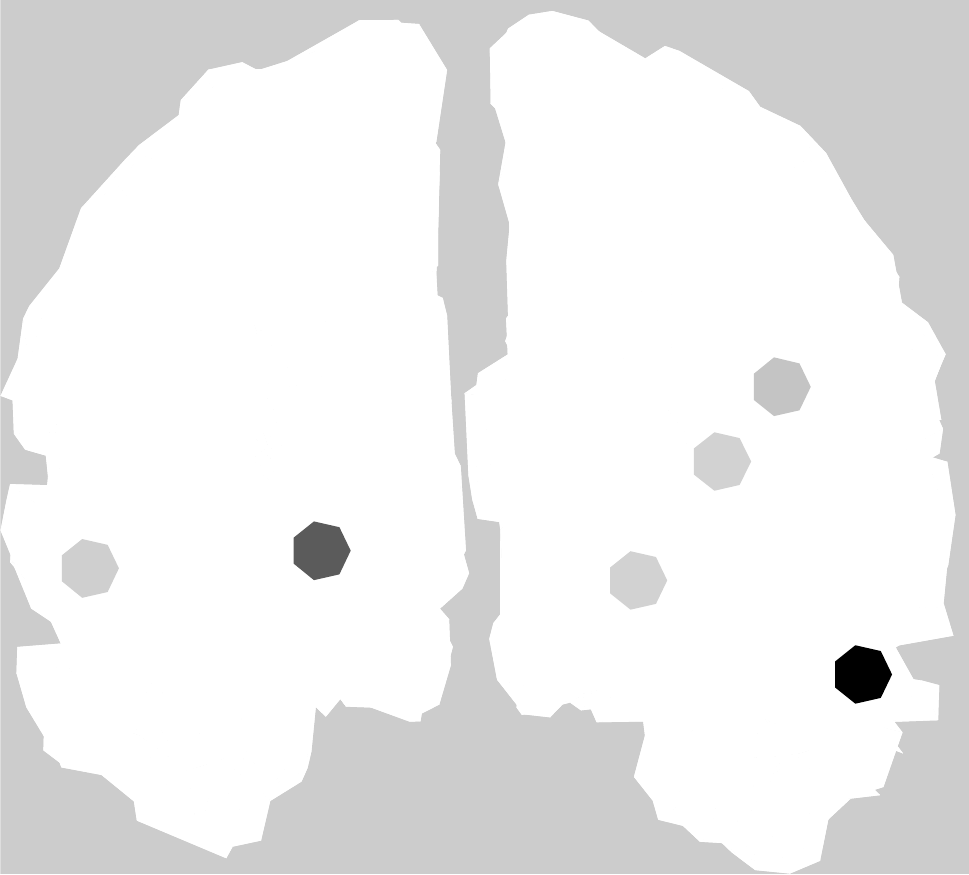}
	}
	
	\subfloat[][Weighted $\ell_{21}$ mixed norm - Manual adjustment of parameter]{
		\includegraphics[width=100pt, height=100pt]{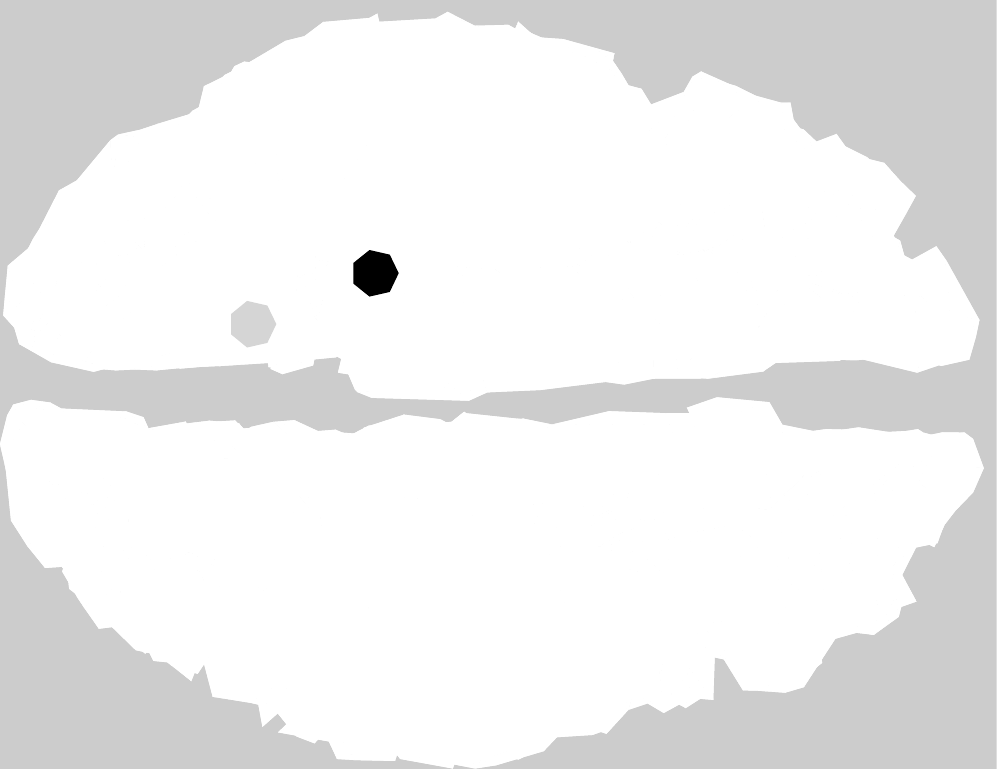}
		\includegraphics[width=100pt, height=100pt]{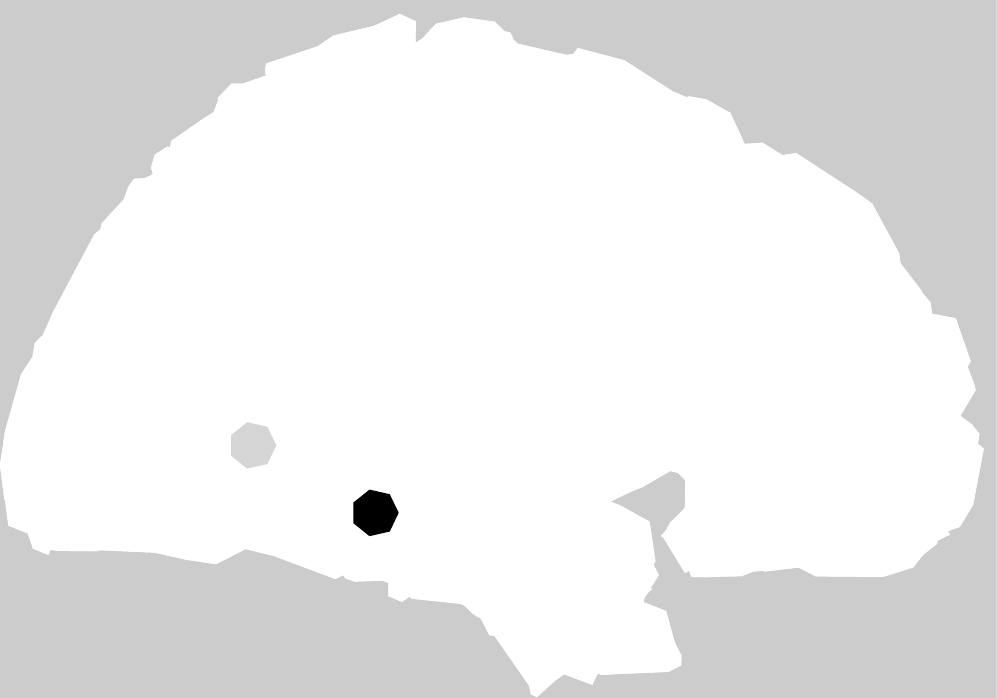}
		\includegraphics[width=100pt, height=100pt]{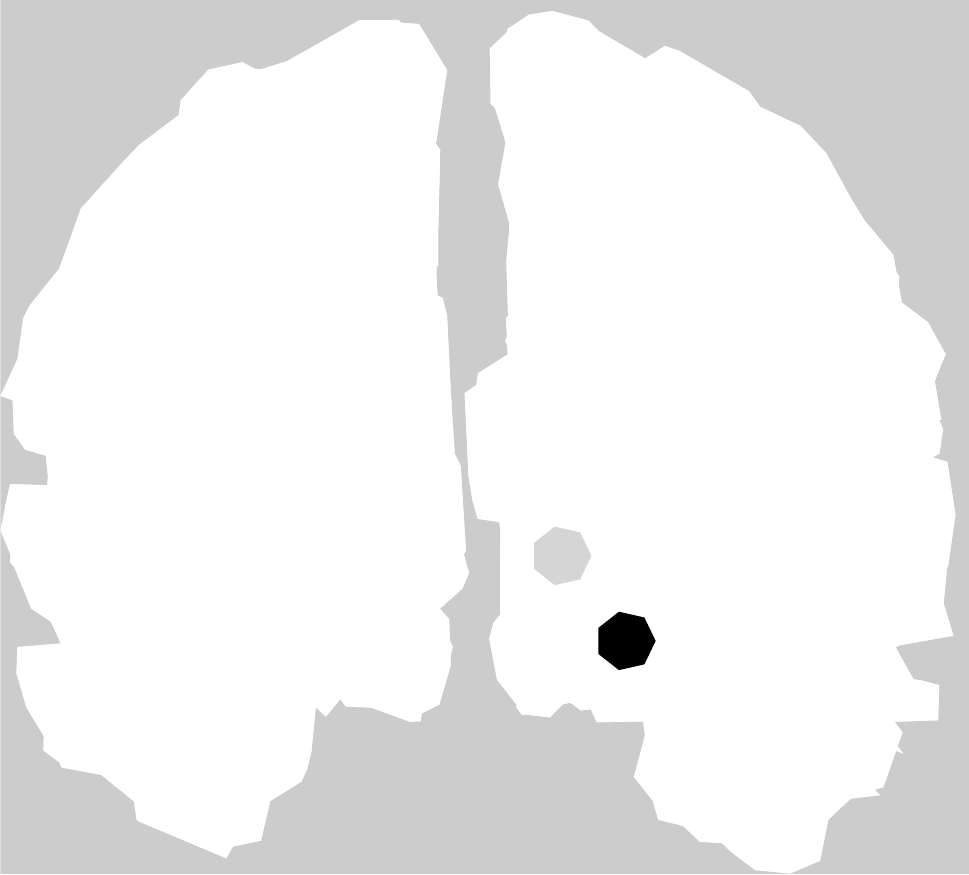}
	}
	
	\centering
	\subfloat[][Proposed method]{
		\includegraphics[width=100pt, height=100pt]{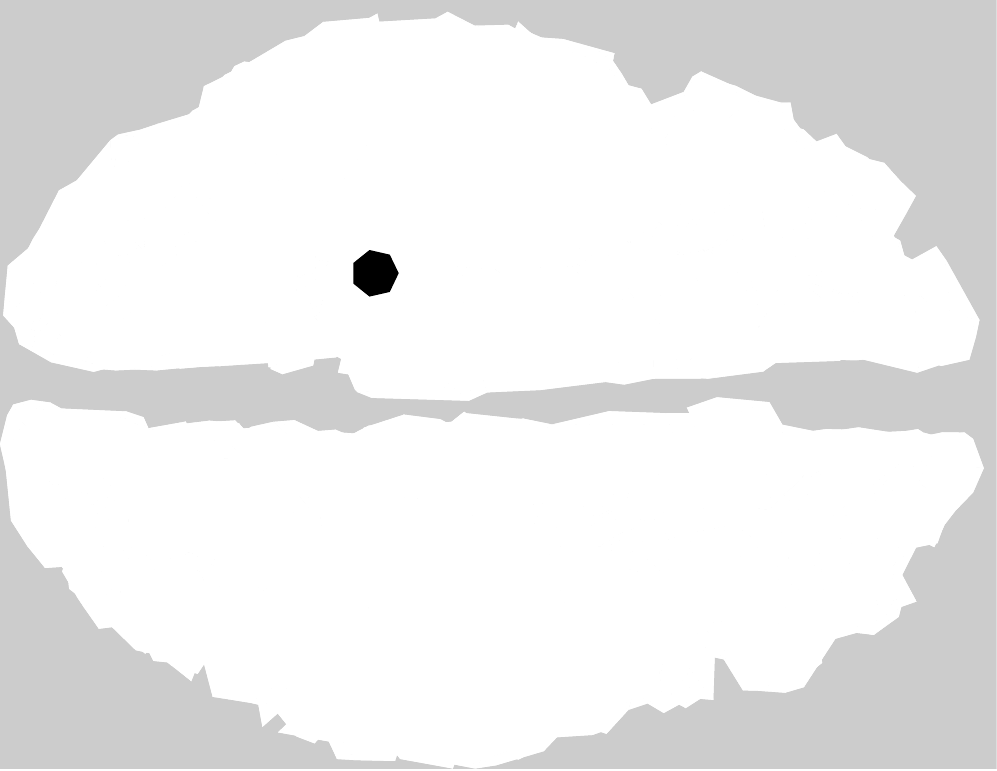}
		\includegraphics[width=100pt, height=100pt]{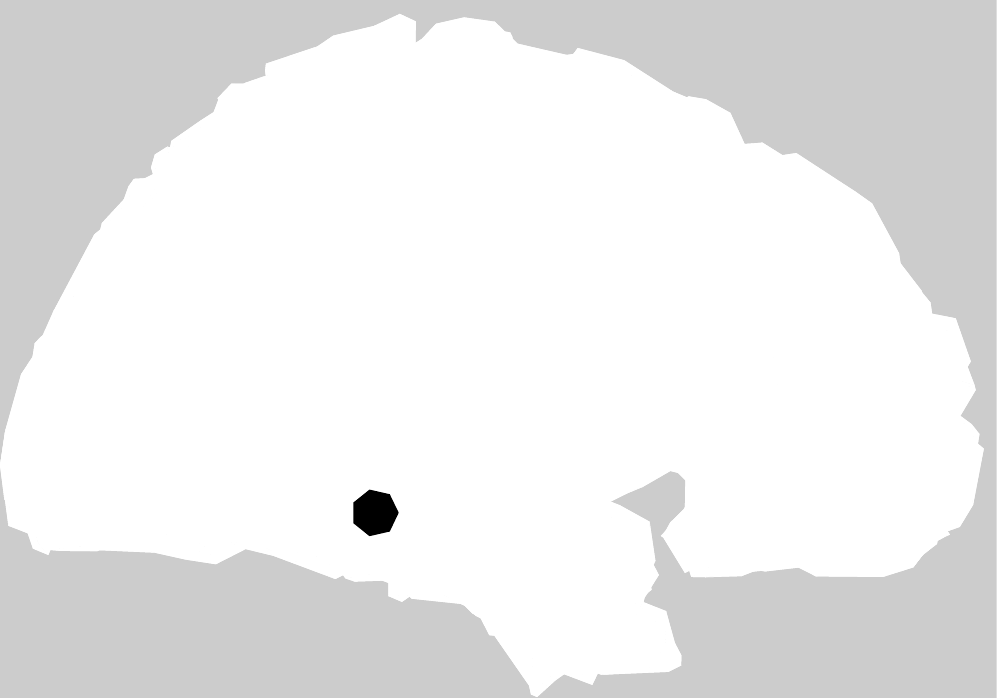}
		\includegraphics[width=100pt, height=100pt]{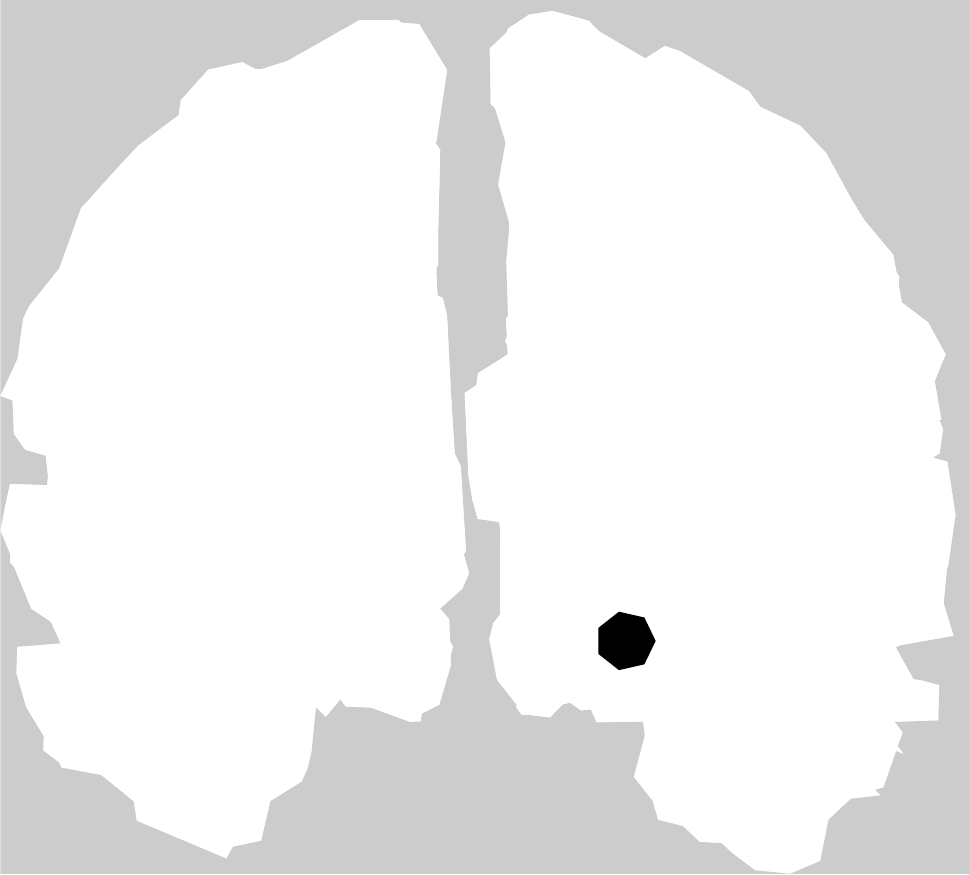}
	}
	
	\centering
	\subfloat[][MSP algorithm]{
		\includegraphics[width=100pt, height=100pt]{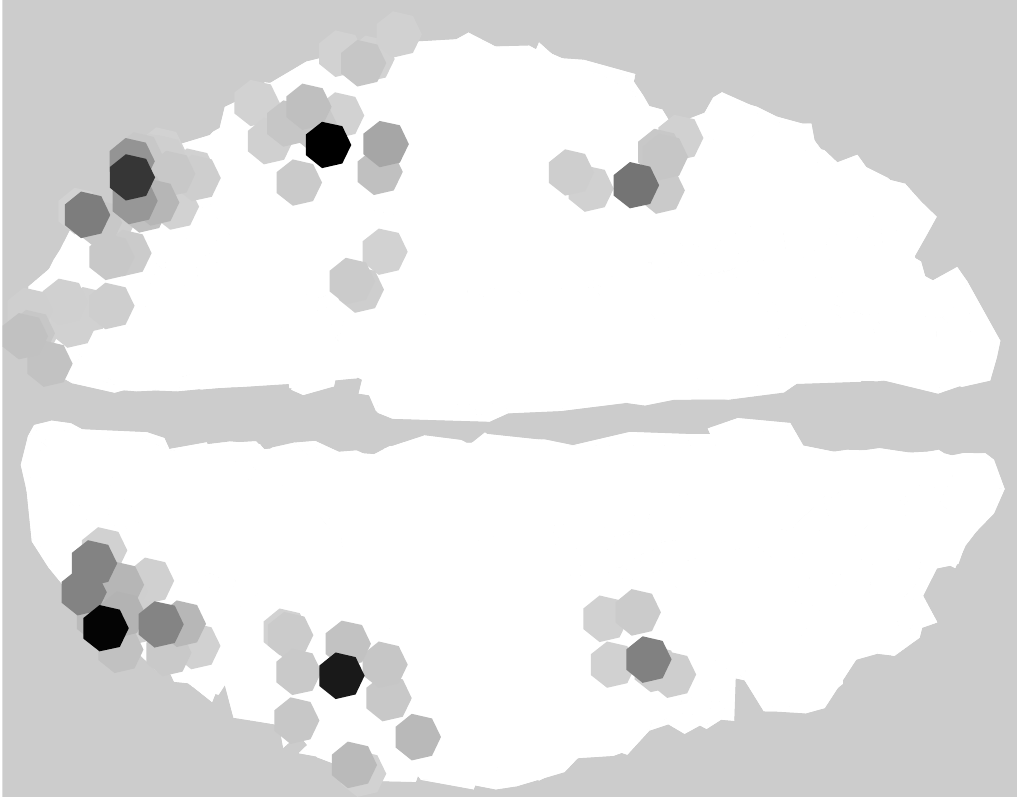}
		\includegraphics[width=100pt, height=100pt]{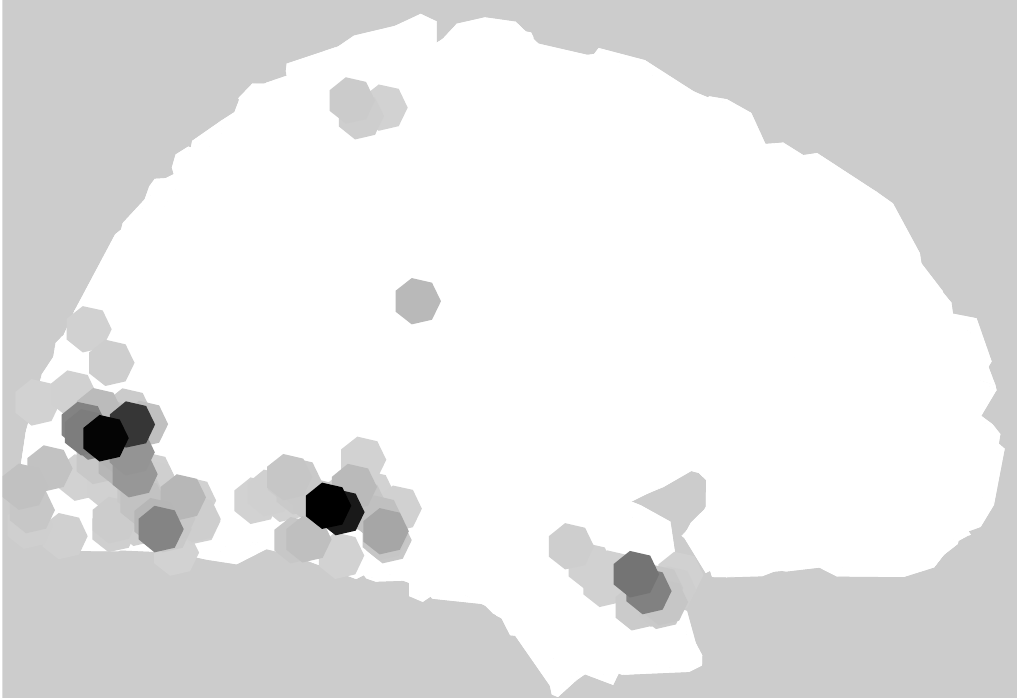}
		\includegraphics[width=100pt, height=100pt]{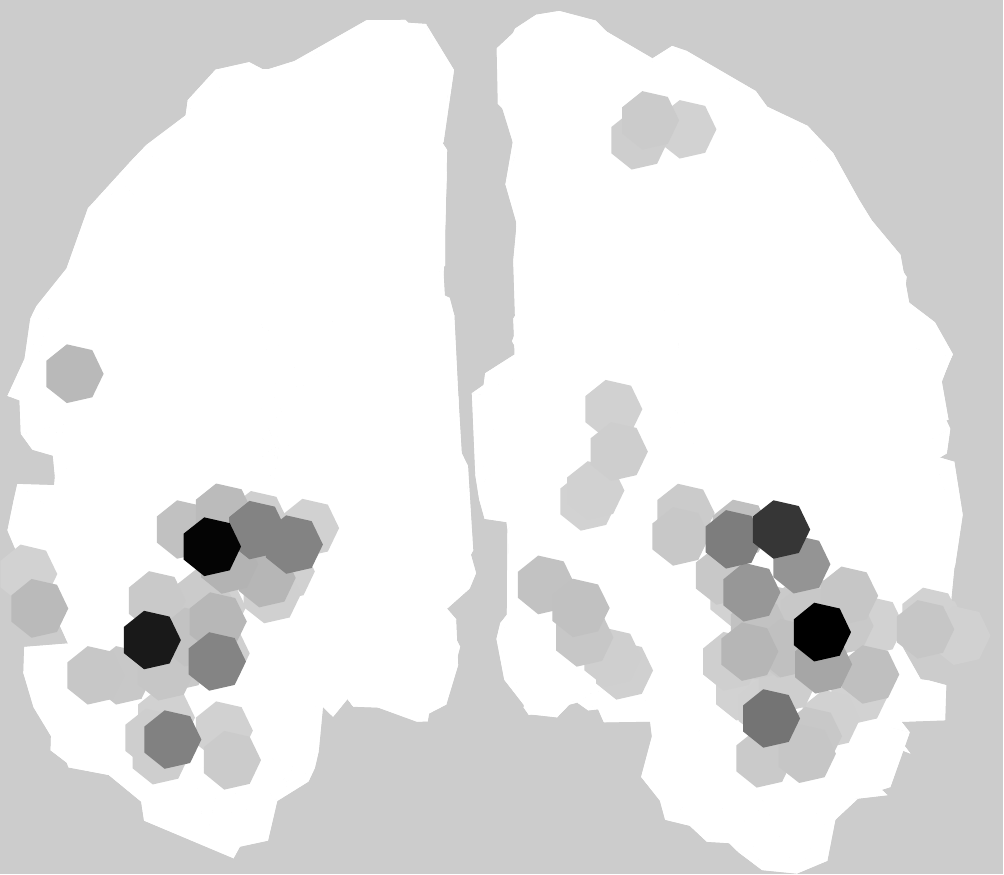}
	}	
		
	\caption{Estimated activity for the facial evoked responses.}
	\label{fig:real_data_location_facial}
\end{figure}

In a second experiment, we used data acquired from a face perception study where the subject was required to evaluate the symmetry of a mixed set of faces and scrambled faces, one of the default datasets of the SPM software\footnote{The SPM software is freely avaiable at http://www.fil.ion.ucl.ac.uk/spm.}. Faces were presented during 600ms every 3600ms. The measurements were taken by the electrodes of a 128-channel ActiveTwo system that sampled at 2048 Hz. The measurements were downsampled to 200Hz and, after artifact rejection, 299 epochs corresponding to the non-scrambled faces were averaged and low-pass filtered to 40Hz. A T1 MRI scan was then downsampled to generate a $3004$ dipole head model.
The estimated activities are shown in Fig. \ref{fig:real_data_location_facial}. As in the previous case, we can see that the $\ell_{21}$ mixed norm response (with the regularization parameter adjusted according to the uncertainty principle) estimates the activity spread around the brain. In contrast, adjusting its regularization parameter manually results in a focal response concentrated in one of the fusiform regions in the temporal lobe associated with the facial recognition process \cite{kanwisher1997fusiform} similar to the one obtained by our algorithm. The MSP algorithm spreads the activity over brain regions located more to the lateral and posterior parts of the brain, further away from the expected area.

Fig. \ref{fig:real_data_waveforms_all_facial} shows the EEG measurements and the estimated waveforms by the $\ell_{21}$ approach and the proposed method. As with the auditory evoked responses data, they differ in the scale due to the underestimation of the activity amplitude by the $\ell_{21}$ approach.

\subsection{Computational cost}
It is important to note that the price to pay with the proposed method, while having several advantages over the $\ell_{21}$ mixed norm approach, is its higher computational complexity as is typical of MCMC methods when compared with optimization techniques. The low SNR three-dipole experiment was processed in $6$ seconds running in a modern Xeon CPU E3-1240 @ 3.4GHz processor (using a Matlab implementation with MEX files written in C) against $104$ms for the $\ell_{21}$ mixed norm approach. However, it is interesting to mention that the $\ell_{21}$ norm approach requires running the algorithm multiple times to adjust the regularization parameter by cross-validation.

\section{Conclusion}
\label{sec:conclusion}
We presented a Bayesian mathematical model for sparse EEG reconstruction that approximates the $\ell_{20}$ mixed norm in a Bayesian framework by a multivariate Bernoulli Laplacian prior. A partially collapsed Gibbs sampler was used to sample from the target posterior distribution. We introduced multiple dipole shift proposals within each MCMC chain and exchange moves between different chains to improve the convergence speed. Using the generated samples, the source activity was estimated jointly with the model hyperparameters in an unsupervised framework. The proposed method was compared with the $\ell_{21}$ mixed norm and the multiple sparse prior methods for a wide variety of situations including several multidipole synthetic activations and two different sets of real data. Our algorithm presented several advantages including better recovery of dipole locations and waveforms in low SNR conditions, the capacity of correctly detecting a higher amount of non-zeros, providing sparser solutions and avoiding underestimation of the activation amplitude. Finally, the possibility of providing several solutions with their corresponding probabilities is interesting. Future work will be devoted to a generalization of the proposed model to cases where the head model is not precisely known.

\newpage
\appendix
\section{Derivation of the conditional probability distributions}
\label{appendix:cond_dist_derivation}
We will now proceed to show the derivation of the conditional probability distributions of the associated model presented in section \ref{sec:bayesian_model} of the report in detail.

\subsection{Posterior distribution}
As specified in the report, the associated posterior distribution is:
\begin{equation}
f(\Yb, \sigma_n^2, \Xb, \zb, a, \taub^2, \omega) = f(\Yb | \Xb, \sigma_n^2) f(\Xb | \taub^2, \zb, \sigma_n^2) f(\zb | \omega) f(\taub^2 | a) f(\sigma_n^2) f(a) f(\omega)
\end{equation}

From it we can derive the conditional distributions of all the associated parameters and hyperparameters using Bayes' theorem:
\begin{align}
f(z_i, \xb_i | \Yb, \Xb_{-i}, \sigma_n^2, \tau_i^2, \omega) &\propto f(\Yb | \Xb, \sigma_n^2) f(\xb_i | \tau_i^2, z_i, \sigma_n^2) f(z_i | \omega)\\
f(\tau_i^2 | \xb_i, \sigma_n^2, a, z_i) &\propto f(\xb_i | \tau_i^2, z_i, \sigma_n^2) f(\tau_i^2 | a)\\
f(a | \taub^2) &\propto f(\taub^2 | a) f(a)\\
f(\sigma_n^2 | \Yb, \Xb, \taub^2, \zb) &\propto f(\Yb | \Xb, \sigma_n^2) f(\Xb | \taub^2, \zb, \sigma_n^2) f(\sigma_n^2)\\
f(\omega | \zb) &\propto f(\zb | \omega) f(\omega)
\end{align}

\subsection{Conditional distributions}
\underline{Conditional distribution of $\tau_i^2$}\\
The conditional distribution of $\tau_i^2$ is
 
\begin{equation}
f(\tau_i^2 | \xb_i, \sigma_n^2, a, z_i) \propto f(\xb_i | \tau_i^2, z_i, \sigma_n^2) f(\tau_i^2 | a)
\end{equation}

that is equal to

\begin{equation}
f(\tau_i^2 | \xb_i, \sigma_n^2, a, z_i) =
\left\{
	\begin{array}{ll}
		\delta(\xb_i) \mathcal{G}\Big(\tau_i^2 \Big| \frac{T + 1} {2}, \frac{v_i a} {2}\Big) & \mbox{if } z_i = 0 \\
		\mathcal{N}\Big(\xb_i \Big| 0, \sigma_n^2 \tau_i^2 I_T\Big) \mathcal{G}\Big(\tau_i^2 \Big| \frac{T + 1} {2}, \frac{v_i a} {2}\Big) & \mbox{if } z_i = 1.
	\end{array}
\right.
\end{equation}

Based on the development of \cite{raman2009bayesian} it can be seen that the conditional distribution of $\tau_i^2$ is a generalized inverse gaussian when $z_i = 1$ and a gamma distribution when $z_i = 0$

\begin{equation}
f(\tau_i^2 | \xb_i, \sigma_n^2, a, z_i) =
\left\{
	\begin{array}{ll}
	\mathcal{G}\Big(\tau_i^2 \Big| \frac{T + 1} {2}, \frac{v_i a} {2}\Big) & \mbox{if } z_i = 0 \\
	\mathcal{GIG}\Big(\tau_i^2 \Big| \frac{1} {2},  v_i a, \frac{||\xb_i||^2} {\sigma_n^2}\Big) & \mbox{if } z_i = 1.\\
	\end{array}
\right.
\end{equation}

\underline{Conditional distribution of $z_i$ and $\xb_i$}\\
As stated in the report, we will sample $z_i$ and $\xb_i$ jointly from

\begin{equation}
f(z_i, \xb_i | \Yb, \Xb_{-i}, \sigma_n^2, \tau_i^2, \omega) \propto f(\Yb | \Xb, \sigma_n^2) f(\xb_i | \tau_i^2, z_i, \sigma_n^2) f(z_i | \omega)
\end{equation}

that is equal to

\begin{align}
\label{eq:l20_cond_xi_init}
&f(z_i, \xb_i | \Yb, \Xb_{-i}, \sigma_n^2, \tau_i^2, \omega) \propto \nonumber \\ &\exp\Big(-\frac{||\Hb \Xb - \Yb||^2} {2 \sigma_n^2}\Big) \Big[(1-z_i) \delta(\xb_i) + z_i \mathcal{N}\Big(\xb_i \Big| 0, \sigma_n^2 \tau_i^2 I_T\Big)\Big] \Big[(1-\omega)\delta(z_i) + \omega \delta(z_i - 1)\Big] = \nonumber \\
&\exp\Big(-\frac{||\Hb \Xb - \Yb||^2} {2 \sigma_n^2}\Big) \Big[(1-\omega) \delta(z_i) \delta(\xb_i) + \omega \delta(z_i - 1) \mathcal{N}\Big(\xb_i \Big| 0, \sigma_n^2 \tau_i^2 I_T\Big)\Big] = \nonumber \\ &(1 - \omega) \delta(z_i) \exp\Big(-\frac{||\Hb \Xb_{-i} - \Yb||^2} {2 \sigma_n^2}\Big) \delta(\xb_i) + \omega ({2 \pi \tau_i^2 \sigma_n^2})^{-\frac{T} {2}} \delta(z_i - 1) \exp\Big(-\frac{||\Hb \Xb - \Yb||^2} {2 \sigma_n^2}\Big) \exp\Big(-\frac{||\xb_i||^2} {2 \sigma_n^2 \tau_i^2}\Big).
\end{align}

The marginal distribution of $z_i$ is of the form
\begin{equation}
\label{eq:l20_cond_form}
f(z_i | \Yb, \Xb_{-i}, \sigma_n^2, \tau_i^2, \omega) = \int f(z_i, \xb_i | \Yb, \Xb_{-i}, \sigma_n^2, \tau_i^2, \omega) d{\xb_i} \propto k_0 \delta(z_i) + k_1 \delta(z_i - 1)
\end{equation}
with

\begin{align}
\label{eq:l20_cond_xi_k0}
&k_0 = \int {(1 - \omega) \exp\Big(-\frac{||\Hb \Xb_{-i} - \Yb||^2} {2 \sigma_n^2}\Big) \delta(\xb_i)} d{\xb_i} = (1 - \omega) \exp\Big(-\frac{||\Hb \Xb_{-i} - \Yb||^2} {2 \sigma_n^2}\Big)\\
\label{eq:l20_cond_xi_k1}
&k_1 = \omega ({2 \pi \tau_i^2 \sigma_n^2})^{-\frac{T} {2}} \int {\exp\Big[-\frac{1}{2 \sigma_n^2} \Big(||\Hb \Xb - \Yb||^2 + \frac{||\xb_i||^2} {\tau_i^2}\Big)\Big]} d{\xb_i}.
\end{align}

This implies that $z_i$ has the following Bernoulli distribution

\begin{equation}
f(z_i | \Yb, \Xb_{-i}, \sigma_n^2, \tau_i^2, \omega) = \mathcal{B} \Big(z_i \Big| \frac{k_1} {k_0 + k_1}\Big).
\end{equation}

To find the value of $k_1$ we calculate the minus logarithm of the integrand
\begin{equation}
\label{eq:l20_cond_xi_minus_log}
-\log\Big(\exp\Big[-\frac{1}{2 \sigma_n^2} \Big(||\Hb \Xb - \Yb||^2 + \frac{||\xb_i||^2} {\tau_i^2}\Big)\Big]\Big) = \frac{1} {2 \sigma_n^2} \Big(||\Hb \Xb - \Yb||^2 + \frac{||\xb_i||^2} {\tau_i^2}\Big)
\end{equation}

and express it as a sum for the different values of $t$
\begin{equation}
\frac{1} {2 \sigma_n^2} \Big(||\Hb \Xb - \Yb||^2 + \frac{||\xb_i||^2} {\tau_i^2}\Big) =
\frac{1} {2 \sigma_n^2} \sum_{t=1}^T \Big(||\Hb \xb^t - \yb^t||^2 + \frac{({\xb_i^t})^2} {\tau_i^2}\Big).
\end{equation}
If we denote $\hb^i$ each column of the operator $\Hb$ we can express term number $t$ of the sum as

\begin{equation}
\label{eq:l20_cond_xi_sep_t}
\frac{1} {2 \sigma_n^2} \Big(||\Hb \xb^t - \yb^t||^2 + \frac{({x_i^t})^2} {\tau_i^2}\Big) = \frac{1} {2 \sigma_n^2} \Big[\Big(\sum_{j \neq i} \hb^j x_j^t + \hb^i x_i^t - \yb^t\Big)^T \Big(\sum_{j \neq i} \hb^j x_j^t + \hb^i x_i^t - \yb^t \Big) + \frac{({x_i^t})^2} {\tau_i^2}\Big].
\end{equation}
By denoting $\Db_i^t = \Yb^t - \sum_{j \neq i} \hb^j x_j^t$ and expanding this expression we have

\begin{equation}
\eqref{eq:l20_cond_xi_sep_t} = \frac{1} {2 \sigma_n^2} \Big(({\hb^i})^T \hb^i ({x_i^t})^2 - 2 x_i^t ({\hb^i})^T \Db_i^t + ({\Db_i^t})^T \Db_i^t + \frac{({x_i^t})^2} {\tau_i^2}\Big).
\end{equation}

Matching the terms of the previous expression with $\frac{(x_i^t-\mu_i^t)^2} {2 \sigma_i^2} + K_i^t$ we have

\begin{align}
&\frac{1} {2 \sigma_n^2} \Big({(\hb^i})^T {\hb^i} ({x_i^t})^2 - 2 x_i^t ({\hb^i})^T \Db_i^t + ({\Db_i^t})^T \Db_i^t + \frac{({x_i^t})^2} {\tau_i^2}\Big) = \frac{({x_i^t})^2} {2 \sigma_i^2} - \frac{\mu_i^t x_i^t} {\sigma_i^2} + \frac{({\mu_i^t})^2} {2 \sigma_i^2} + K_i^t\\
&\frac{1} {2 \sigma_i^2} = \frac{({\hb^i})^T {\hb^i} + \frac{1} {\tau_i^2}} {2 \sigma_n^2} => \sigma_i^2 = \frac{\sigma_n^2 \tau_i^2} {1 + \tau_i^2 ({\hb^i})^T \hb^i}\\
&\frac{\mu_i^t} {\sigma_i^2} = \frac{({\hb^i})^T \Db_i^t} {\sigma_n^2} => \mu_i^t = \frac{\sigma_i^2 ({\hb^i})^T \Db_i^t} {\sigma_n^2}\\
&\frac{({\mu_i^t})^2} {2 \sigma_i^2} + K_i^t = \frac{({\Db_i^t})^T \Db_i^t} {2 \sigma_n^2} => K_i^t = \frac{({\Db_i^t})^T \Db_i^t} {2 \sigma_n^2} - \frac{({\mu_i^t})^2} {2 \sigma_i^2}.
\end{align}

Summing over all time samples and applying the function $\exp(-x)$ (to compensate the steps done in \ref{eq:l20_cond_xi_minus_log} and \ref{eq:l20_cond_xi_sep_t}) results in
\begin{equation}
\label{eq:l20_cond_xi_last_step}
\exp\Big(-\frac{||\Hb \Xb - \Yb||^2} {2 \sigma_n^2}\Big) \exp\Big(-\frac{||\xb_i||^2} {2 \sigma_n^2 \tau_i^2}\Big) = ({2 \pi \sigma_i})^{\frac{T} {2}} \prod_{t=1}^T {\exp(-K_i^t) \mathcal{N}\Big(x_i^t \Big| \mu_i^t, \sigma_i^2\Big)}.
\end{equation}

We can calculate the final value of $k_1$ by combining \eqref{eq:l20_cond_xi_k1} and \eqref{eq:l20_cond_xi_last_step}
\begin{align}
&k_1 = \omega ({2 \pi \tau_i^2 \sigma_n^2})^{-\frac{T} {2}} \int \exp\Big(-\frac{||\Hb \Xb - \Yb||^2} {2 \sigma_n^2}\Big) \exp\Big(-\frac{||\xb_i||^2} {2 \sigma_n^2 \tau_i^2}\Big) d{\xb_i} = \omega{(2 \pi \tau_i^2 \sigma_n^2)}^{-\frac{T} {2}} ({2 \pi \sigma_i})^{\frac{T} {2}} \prod_{t=1}^T \exp(-K_i^t) = \nonumber \\
&\omega{\Big(\frac{\sigma_n^2 \tau_i^2} {\sigma_i^2}\Big)}^{-\frac{T} {2}} \exp\Big(-\frac{||\Hb \Xb_{-i} - \Yb||^2} {2 \sigma_n^2}\Big) \exp\Big(\frac{||\mub_i||^2} {2 \sigma_i^2}\Big). 
\end{align}

Using \eqref{eq:l20_cond_xi_init} and \eqref{eq:l20_cond_xi_last_step} we obtain the conditional distribution of $\xb_i$

\begin{equation}
f(\xb_i | z_i, \Yb, \Xb_{-i}, \sigma_n^2, \tau_i^2) = 
\left\{
	\begin{array}{ll}
		\delta(\xb_i)  & \mbox{if } z_i = 0 \\
		\mathcal{N}\Big(\xb_i \Big| \mub_i, \sigma_i^2 I_T\Big) & \mbox{if } z_i = 1.
	\end{array}
\right.
\end{equation}

\underline{Conditional distribution of $a$}\\
The conditional distribution of $a$ is

\begin{equation}
f(a | \taub^2) \propto f(a) \prod_{i=1}^N{f(\tau_i^2 | \xb_i, \sigma_n^2, a)}
\end{equation} 

\begin{equation}
f(a | \taub^2) \propto a^{\alpha-1} \exp\Big(-\beta a\Big) \prod_{i = 1}^N \Big[{{\Big(\frac{a v_i} {2}\Big)}^{\frac{T+1} {2}} {(\tau_i^2)}^{\frac{T - 1} {2}} \exp\Big(-\frac {a v_i \tau_i^2} {2}\Big)\Big]}\propto a^{\alpha + \frac{N(T+1)} {2} - 1} \exp\Big(-\Big(\beta + \frac{1} {2} \sum_{i=1}^N[v_i \tau_i^2]\Big)a \Big)
\end{equation} 

which corresponds to the following gamma distribution

\begin{equation}
f(a | \taub^2) = \mathcal{G}\Big(a \Big| \frac{N(T+1)} {2} + \alpha, \frac{\sum_i[v_i \tau_i^2]} {2} + \beta\Big).
\end{equation}

\underline{Conditional distribution of $\sigma_n^2$}\\
The conditional distribution of $\sigma_n^2$ is

\begin{equation}
f(\sigma_n^2 | \Yb, \Xb, \taub^2, \zb) \propto f(\Yb | \Xb, \sigma_n^2) f(\Xb | \taub^2, \zb, \sigma_n^2) f(\sigma_n^2)
\end{equation}

\begin{align}
&f(\sigma_n^2 | \Yb, \Xb, \taub^2, \zb) \propto (\sigma_n^2)^{-1} {(2 \pi \sigma_n^2)}^{-\frac{MT} {2}} \exp\Big(-\frac{||\Hb \Xb - \Yb||^2} {2 \sigma_n^2}\Big) \nonumber \prod_{i=1}^N{\Big[{(z_i - 1) \delta(\xb_i) + z_i \mathcal{N}\Big(\xb_i \Big| 0, \sigma_n^2 \tau_i^2 I_T\Big)}\Big]}.
\end{align}

Denoting $I_k = \{i : z_i = k\}$ for $k = \{0, 1\}$ and using the identity

\begin{equation}
\label{eq:acc_prob_prod_indentity}
\prod_{i=1}^N \Big[{(z_i - 1) f(x) + z_i g(x)}\Big] = {\prod_{i \in I_0} f(x)} {\prod_{i \in I_1} g(x)}
\end{equation}

we can express this by

\begin{align}
f(\sigma_n^2 | \Yb, \Xb, \taub^2, \zb) \propto &(\sigma_n^2)^{-1} {(2 \pi \sigma_n^2)}^{-\frac{MT} {2}} \exp\Big(-\frac{||\Hb \Xb - \Yb||^2} {2 \sigma_n^2}\Big) \prod_{i \in I_0}{\delta(\xb_i)} \prod_{i \in I_1} \Big[{{(2 \pi \tau_i^2 \sigma_n^2)}^{-\frac{T} {2}} \exp\Big(-\frac{||\xb_i||^2} {2 \tau_i^2 \sigma_n^2}\Big)}\Big].
\end{align}

The previous expression leads to
\begin{equation}
f(\sigma_n^2 | \Yb, \Xb, \taub^2, \zb) \propto (\sigma_n^2)^{-\Big(1 + \frac{(M+||\zb||_0)T} {2}\Big)} \exp\Big(-\frac{1} {2 \sigma_n^2} \Big[||\Hb \Xb - \Yb||^2 + \sum_{i\ \in I_1} {\frac{||\xb_i||^2} {\tau_i^2}}\Big]\Big)
\end{equation}

which corresponds to the following inverse gamma distribution

\begin{equation}
f(\sigma_n^2 | \Yb, \Xb, \taub^2, \zb) = \mathcal{IG}\Big(\sigma_n^2 \Big| \frac{(M+||\zb||_0)T} {2}, \frac{1} {2} \Big[||\Hb \Xb - \Yb||^2 + \sum_{i \in I_1} \frac{||\xb_i||^2} { \tau_i^2}\Big]\Big).
\end{equation}

\underline{Conditional distribution of $\omega$}\\
The conditional distribution for $\omega$ is

\begin{equation}
f(\omega | \zb) \propto f(\zb | \omega) f(\omega)
\end{equation}

\begin{equation}
f(\omega | \zb) \propto 1_{0,1} \prod_{i=1}^{N} \Big[\delta(z_i) (1 - \omega) + \delta(z_i - 1) \omega \Big]
\end{equation}

where $1_{0,1}$ represents a function that is 1 for $0 < \omega < 1$ and 0 elsewhere. Using the identity \eqref{eq:acc_prob_prod_indentity} this can be shown to be equal to

\begin{equation}
f(\omega | \zb) \propto 1_{0,1} (1 - \omega)^{N - ||\zb||_0}{\omega}^{||\zb||_0}\prod_{i \in I_0}{\delta(z_i)} \prod_{i \in I_1} {\delta(z_i - 1)}
\end{equation}

which corresponds to the following Beta distribution

\begin{equation}
f(\omega | \zb) = \mathcal{B}e\Big(\omega \Big| 1 + ||\zb||_0, 1 + N - ||\zb||_0\Big).
\end{equation}

\newpage
\section{Derivation of proposal acceptance probability}
\label{appendix:accept_proposal_prob}
In section \ref{sec:convergence_considerations} we presented two different kind of proposals used in the algorithm: multiple dipole shifts and inter-chain proposals. Both of them consists in proposing to change the values of the vectors $\zb$ and $\taub^2$ jointly. In this appendix we will calculate their acceptance probability.

Using Bayes' theorem it is easy to show that the acceptance probability is equal to the ratio of the conditional distribution $f(\taub^2, \zb | \Yb, a, \sigma_n^2, \omega)$ evaluated in the current and proposed values of the vectors. It is possible to improve the speed of the calculation by only considering the $i$th row if the current value of the element $z_i$ differs from the proposed value $z^p_i$. Denoting this subset of rows by $\rb$ we can calculate $f(\taub^2_{\rb}, \zb_{\rb}, \Xb_{\rb} | \Yb, \Xb_{-r}, \zb_{-r}, \taub_{-r}^2, a, \sigma_n^2, \omega)$ and then integrate it with respect to $\Xb_r$.

Based on the posterior \eqref{eq:posterior}, we can derive the conditional distribution of the subset $\rb$ of rows as
\begin{equation}
f(\taub_r^2, \zb_r, \Xb_r | \Yb, \Xb_{-r}, \zb_{-r}, \taub_{-r}^2, a, \sigma_n^2, \omega) \propto f(\Yb | \Xb, \sigma_n^2) \prod_{i \in \rb} {f(\xb_i, z_i | \sigma_n^2, \tau_i^2, \omega)} \prod_{i \in \rb} {f(\tau_i^2 | a)}
\end{equation}
which implies
\begin{align}
f(\taub_r^2, \zb_r, \Xb_r | .) \propto &\prod_{t=1}^T {\mathcal{N}\Big(\yb^t \Big| \Hb \xb^t, \sigma_n^2 I_M\Big)}\\ &\prod_{i \in \rb} {\Big[(1-\omega) \delta(z_i) \delta(\xb_i) + \omega \delta(z_i - 1) \mathcal{N}\Big(\xb_i \Big| 0, \sigma_n^2 \tau_i^2 I_T\Big)\Big]} \nonumber \\ &\prod_{i \in \rb}{\mathcal{G}\Big(\tau_i^2 \Big| \frac{T + 1} {2}, \frac{v_i a} {2}\Big)}. \nonumber
\end{align}

Denoting the subsets: $\Ib_k = \{r_i : z_{\rb_i} = k\}$ (Note that $\Ib_0 \cup \Ib_1 = \rb$ and $\Ib_0 \cap \Ib_1 = \emptyset$) and their cardinals $C_k = \#\Ib_k$ for $k = \{0, 1\}$ and using identity \eqref{eq:acc_prob_prod_indentity} allows us to express this as

\begin{align}
f(\taub_r^2, \zb_r, \Xb_r | .) \propto &\prod_{t=1}^T \mathcal{N}\Big(\yb^t \Big| \Hb \xb^t, \sigma_n^2 I_M\Big)\\ &\Big[(1-\omega)^{C_0} \omega^{C_1} \Big(\prod_{i \in \Ib_0} \delta(\xb_i)\Big) \Big(\prod_{i \in \Ib_1} \mathcal{N}(\xb_i | 0, \sigma_n^2 \tau_i^2 I_T)\Big)\Big]\nonumber \\ &\prod_{i \in \rb}{\mathcal{G}\Big(\tau_i^2 \Big| \frac{T + 1} {2}, \frac{v_i a} {2}\Big)}. \nonumber
\end{align}

We now integrate the above expression over $\Xb_r$ leading to

\begin{align}
f(\taub_r^2, \zb_r | .) \propto & \int_{\Xb_r} {f(\taub_r^2, \zb_r, \Xb_r | .)} d{\Xb_r} = (1-\omega)^{C_0} \omega^{C_1} \prod_{i \in \rb}{\mathcal{G}\Big(\tau_i^2 \Big| \frac{T + 1} {2}, \frac{v_i a} {2}\Big)}\\
&\int_{\Xb_r} \Big[\Big({\prod_{t=1}^T \mathcal{N}\Big(\yb^t \Big| \Hb \xb^t, \sigma_n^2 I_M\Big)\Big) \Big(\prod_{i \in \Ib_0} \delta(\xb_i)\Big) \Big(\prod_{i \in \Ib_1} \mathcal{N}\Big(\xb_i | 0, \sigma_n^2 \tau_i^2 I_T\Big)\Big)\Big]} d{\Xb_r}. \nonumber
\end{align}

Evaluating $\xb_i = 0$ for $i \in \Ib_0$ because of the $\delta(\xb_i)$ converts this to

\begin{equation}
\label{eq:acc_prob_integral_with_i}
f(\taub_r^2, \zb_r | .) \propto (1-\omega)^{C_0} \omega^{C_1} I \prod_{i \in \rb}{\mathcal{G}\Big(\tau_i^2 \Big| \frac{T + 1} {2}, \frac{v_i a} {2}\Big)}\\ 
\end{equation}
with 
\begin{equation}
I = \int_{\Xb_{\Ib_1}} \Big[\Big({\prod_{t=1}^T \mathcal{N}\Big(\yb^t \Big| \Hb_{-\Ib_0} \xb_{-\Ib_0}^t, \sigma_n^2 I_M\Big)\Big) \Big(\prod_{i \in \Ib_1} \mathcal{N}\Big(\xb_{i} | 0, \sigma_n^2 \tau_i^2 I_T\Big)\Big)\Big]} d{\Xb_{\Ib_1}}
\end{equation}
being $\xb_{-\sb}^t$ the vector $\xb^t$ that has all the rows in $\sb$ eliminated and $\Hb_{-\sb}$ the matrix $\Hb$ that has all the columns in $\sb$ eliminated.

$I$ can also be expressed as
\begin{align}
I = (2 \pi \sigma_n^2)^{-\frac{M T} {2}} (2 \pi \sigma_n^2)^{-\frac{T C_1} {2}} \prod_{i \in \Ib_1} (\tau_i^2) \int_{\Xb_{\Ib_1}} \exp\Big[{-\frac{1} {2} \sum_{t=1}^T C^t\Big] d{\xb_{\Ib_1}}}
\end{align}
with $C^t = \frac{1} {\sigma_n^2} \Big({||\Hb_{-\Ib_0} \xb_{-\Ib_0}^t - \yb^t||}^2 + \sum_{i \in \Ib_1} {\frac{({x_i^t})^2} {\tau_i^2}}\Big) = \frac{1} {\sigma_n^2} \Big({||\Hb_{-\rb} \xb_{-\rb}^t + \Hb_{\Ib_1} \xb_{\Ib_1}^t - \yb^t||}^2 + \sum_{i \in \Ib_1} {\frac{({x_i^t})^2} {\tau_i^2}}\Big)$.

Denoting $\Db^t = \Hb_{-\Rb} \xb_{-\Rb}^t - \yb^t$ this can be expressed as

\begin{align}
C^t = \frac{1} {\sigma_n^2} \Big({{||\Db^t + \Hb_{\Ib_1} \xb_{\Ib_1}^t||}^2 + \sum_{i \in \Ib_1} {\frac{({x_i^t})^2} {\tau_i^2}}}\Big).
\end{align}

If we denote $\Qb = \diag(\frac{1} {\tau^2_{\rb}})$ being $\diag(\sb)$ the diagonal square matrix formed by the elements of vector $\sb$

\begin{align}
C^t = \frac{1} {\sigma_n^2} \Big({({\xb_{\Ib_1}^t})^T (({\Hb_{\Ib_1})^T \Hb_{\Ib_1} + \Qb}) {\xb_{\Ib_1}^t} + 2 ({\Db^t})^T \Hb_{\Ib_1} ({\xb_{\Ib_1}^t})^T + ({\Db^t})^T \Db^t}\Big).
\end{align}

Matching each term between the previous expression and $\Big(\xb_{\Ib_1}^t - \mu^t\Big)^T \Sigma^{-1} \Big(\xb_{\Ib_1}^t - \mu^t\Big)$

\begin{align}
&{\frac{1} {\sigma_n^2}\Big({({\xb_{\Ib_1}^t})^T ({\Hb_{\Ib_1}^T \Hb_{\Ib_1} + \Qb}) \xb_{\Ib_1}^t + 2 ({\Db^t})^T \Hb_{\Ib_1} ({\xb_{\Ib_1}^t})^T + ({\Db^t})^T {\Db^t}}\Big)} =\\ &({\xb_{\Ib_1}^t})^T \Sigma^{-1} \xb_{\Ib_1}^t - 2 ({\mu^t})^T \Sigma^{-1} ({\xb_{\Ib_1}^t})^T + ({\mu^t})^T  \Sigma^{-1} \mu^t + K^t \nonumber
\end{align}

we get the following equations

\begin{align}
{\frac{1} {\sigma_n^2}\Big({\Hb_{\Ib_1}^T \Hb_{\Ib_1} + \Qb}\Big)} &= \Sigma^{-1}\\
\frac{2 ({\Db^t})^T \Hb_{\Ib_1}} {\sigma_n^2} = - 2 ({\mu^t})^T \Sigma^{-1} => \mu^t &= -\frac{\Sigma \Hb_{\Ib_1}^T \Db^t} {\sigma_n^2} \\
({\mu^t})^T  \Sigma^{-1} \mu^t + K^t = \frac{({\Db^t})^T \Db^t} {\sigma_n^2} => K^t &= \frac{({\Db^t})^T \Db^t} {\sigma_n^2} - ({\mu^t})^T  \Sigma^{-1} \mu^t.
\end{align}

We can now calculate the value of $I$
\begin{align}
I = &(2 \pi \sigma_n^2)^{-\frac{M T} {2}} (2 \pi \sigma_n^2)^{-\frac{T C_1} {2}} \prod_{i \in \Ib_1} (\tau_i^2) \int_{\Xb_{\Ib_1}} \exp\Big[{-\frac{1} {2} \sum_{t=1}^T C^t\Big]} d{\Xb_{\Ib_1}} = \\& (2 \pi \sigma_n^2)^{-\frac{M T} {2}} (2 \pi \sigma_n^2)^{-\frac{T C_1} {2}} \prod_{i \in \Ib_1} (\tau_i^2) \exp\Big({-\frac{\sum_{t=1}^T K^t} {2}}\Big) \int_{\Xb_{\Ib_1}} {\prod_{t=1}^T \exp\Big(-\frac{1} {2} \Big[(x_i^t - \mu_i^t)^T \Sigma^{-1} (x_i^t - \mu_i^t)\Big]\Big)} d{\Xb_{\Ib_1}} =\nonumber \\& (2 \pi \sigma_n^2)^{-\frac{M T} {2}} (\sigma_n^2)^{-\frac{T C_1} {2}} \prod_{i \in \Ib_1} (\tau_i^2) \exp\Big({-\frac{\sum_{t=1}^T K^t} {2}}\Big) |\Sigma|^{\frac{T} {2}}. \nonumber
\end{align}

Finally using the value of $I$ in \eqref{eq:acc_prob_integral_with_i} yields
\begin{align}
&f(\taub_r^2, \zb_r | .) \propto \int_{\xb_r} {f(\taub_r^2, \zb_r, \xb_r | .) d{\xb_r}} \propto \\ & (1-\omega)^{C_0} \omega^{C_1} (\sigma_n^2)^{-\frac{T C_1} {2}} |\Sigma|^{\frac{T} {2}} \prod_{i \in \Ib_1} {(\tau_i^2)^{-\frac{T} {2}}} \exp\Big({-\frac{\sum_{t=1}^T K^t} {2}}\Big) \prod_{i = 1}^{N}{\mathcal{G}\Big(\tau_i^2 \Big| \frac{T + 1} {2}, \frac{v_i a} {2}\Big)}.\nonumber
\end{align}

\footnotesize
\bibliographystyle{IEEEtran}

\end{document}